\documentclass[letter]{aa}
\usepackage{orcidlink}
\usepackage[varg]{txfonts}
\usepackage{svg}
\usepackage{hyperref}
\hypersetup{colorlinks, allcolors=blue}
\usepackage{graphicx,amsmath}
\usepackage{color}
\usepackage{xcolor}
\usepackage{txfonts}
\usepackage{lipsum}
\usepackage{subcaption}
\usepackage{lscape}
\usepackage{placeins}                                
\usepackage{adjustbox}

\newcommand{\Msun}{{\rm\,M_\odot}}
\newcommand{\kpc}{{\rm\,kpc}}
\newcommand{\Gyr}{{\rm\,Gyr}}

\begin{document} 

   \title{ 
   SMUGGLE-Ring: Evolutionary link between nuclear star cluster and nuclear disk}

    \author{SungWon Kwak       \inst{\ref{aip}}\orcidlink{0000-0003-0957-6201},
            Mathias Schultheis \inst{\ref{nice}}\orcidlink{0000-0002-6590-1657},
            Ivan Minchev       \inst{\ref{aip}}\orcidlink{0000-0002-5627-0355},      
            Cristina Chiappini \inst{\ref{aip}}\orcidlink{0000-0003-1269-7282}, \\
            Woong-Tae Kim      \inst{\ref{snu1},\ref{snu2}}\orcidlink{0000-0003-4625-229X}, 
            Seungwon Baek      \inst{\ref{snu1}}\orcidlink{0009-0002-0251-9570},
            Federico Marinacci \inst{\ref{bol1},\ref{bol2}}\orcidlink{0000-0003-3816-7028}, 
            Mark Vogelsberger  \inst{\ref{mit}}\orcidlink{0000-0001-8593-7692}, \\ 
            Laura V. Sales     \inst{\ref{ucr}}\orcidlink{0000-0002-3790-720X},
            Hui Li             \inst{\ref{tsing}}\orcidlink{0000-0002-1253-2763},
            \and 
            Matthias Steinmetz \inst{\ref{aip},\ref{uniP}}\orcidlink{0000-0001-6516-7459}
            }

    \institute{Leibniz-Instit\"ut f\"ur Astrophysik Potsdam (AIP), An der Sternwarte 16, 14482, Potsdam, Germany \label{aip} 
    \email{skwak@aip.de}
    \and Universit\'e C\^ote d'Azur, Observatoire de la C\^ote d'Azur, Laboratoire Lagrange, CNRS, Blvd de l'Observatoire, 06304 Nice, France\label{nice}
    \and Department of Physics \& Astronomy, Seoul National University, Seoul 08826, Republic of Korea\label{snu1}
    \and SNU Astronomy Research Center, Seoul National University, Seoul 08826, Republic of Korea\label{snu2}
    \and Dipartimento di Fisica e Astronomia ``Augusto Righi'', Universit\`a di Bologna, Via Piero Gobetti 93/2, I-40129 Bologna, Italy\label{bol1}
    \and
    INAF, Osservatorio di Astrofisica e Scienza dello Spazio di Bologna, Via Piero Gobetti 93/3, I-40129 Bologna, Italy\label{bol2}
    \and Department of Physics and Kavli Institute for Astrophysics and Space Research, Massachusetts Institute of Technology, Cambridge, MA 02139, USA \label{mit}
    \and University of California, Riverside, 900 University Ave., Riverside, CA 92521, USA \label{ucr}
    \and Department of Astronomy, Tsinghua University, Haidian DS 100084, Beijing, China \label{tsing}
    \and Universit\"at Potsdam, Institut f\"ur Physik und Astronomie, Karl-Liebknecht-Str. 24-25, 14476, Potsdam, Germany \label{uniP}
    }   

   \date{\today}

\abstract{
We present a high-resolution hydrodynamical simulation of the formation and evolution of nuclear structures in a Milky Way-mass galaxy using the SMUGGLE model. The system naturally develops a bar in isolation of $\approx5$ kpc in length, driving sustained gas inflows toward the center that lead to the formation of a nuclear stellar disk (NSD) and a nuclear star cluster (NSC). By only considering stars born after bar formation, we can cleanly isolate the nuclear structures and recover a clear inside-out growth of the NSD. In line with observations, we find that stellar feedback induces repeated shocks that regulate the size of the nuclear gas disk and drive gas from its outer edge toward the NSC region. Over time, the NSD and NSC share similar mass growth and star formation histories, except during the accretion of a massive star cluster. Our results suggest that both the evolutionary timescale of the bar (and thus of the NSD) and the accretion history of star clusters are essential for obtaining tighter scaling relations for nuclear structures and their host galaxies. Finally, our results favor a lower bulge mass for the Milky Way than that of our model ($B/D\approx 0.045$) to explain the compact size of its nuclear disk.}

   \titlerunning{Link between NSCs and NSDs}
   \authorrunning{Kwak et al.}
   \maketitle
   \nolinenumbers 

\section{Introduction}
Stellar bars are common features in disk galaxies \citep{eskridge00, wang25}, which go through  dynamical interactions with both the dark matter (DM) halo and the gaseous medium. Angular momentum exchange with the DM halo promotes bar growth and slowdown \citep{athanassoula03, kwak17, kwak26a}, while the nonaxisymmetric perturbation from the bar drives gas inflows toward the galactic center \citep{combes85, athanassoula92, combes93, kwak26b}, thereby exerting positive torques on the bar rotation \citep{beane23, kwak26b}. Depending on the spheroidal bulge mass and the resulting bar properties (which control the location of the inner Lindblad resonance and the efficiency of gas inflow) the nuclear structures follow markedly different evolutionary paths \citep{kwak26b}.

\begin{figure*}[t!]
    \centering
    \setlength{\tabcolsep}{2pt}
    \begin{tabular}{@{}c@{\hspace{5pt}}c@{\hspace{5pt}}c@{}}

    \multicolumn{3}{c}{\textbf{\large 1.5 Gyr}} \\[-1pt]
    \rotatebox{90}{\large \textbf{y [kpc]}}\hspace{-3pt} &
    \parbox[c]{0.34\textwidth}{\centering\includegraphics[width=0.34\textwidth]{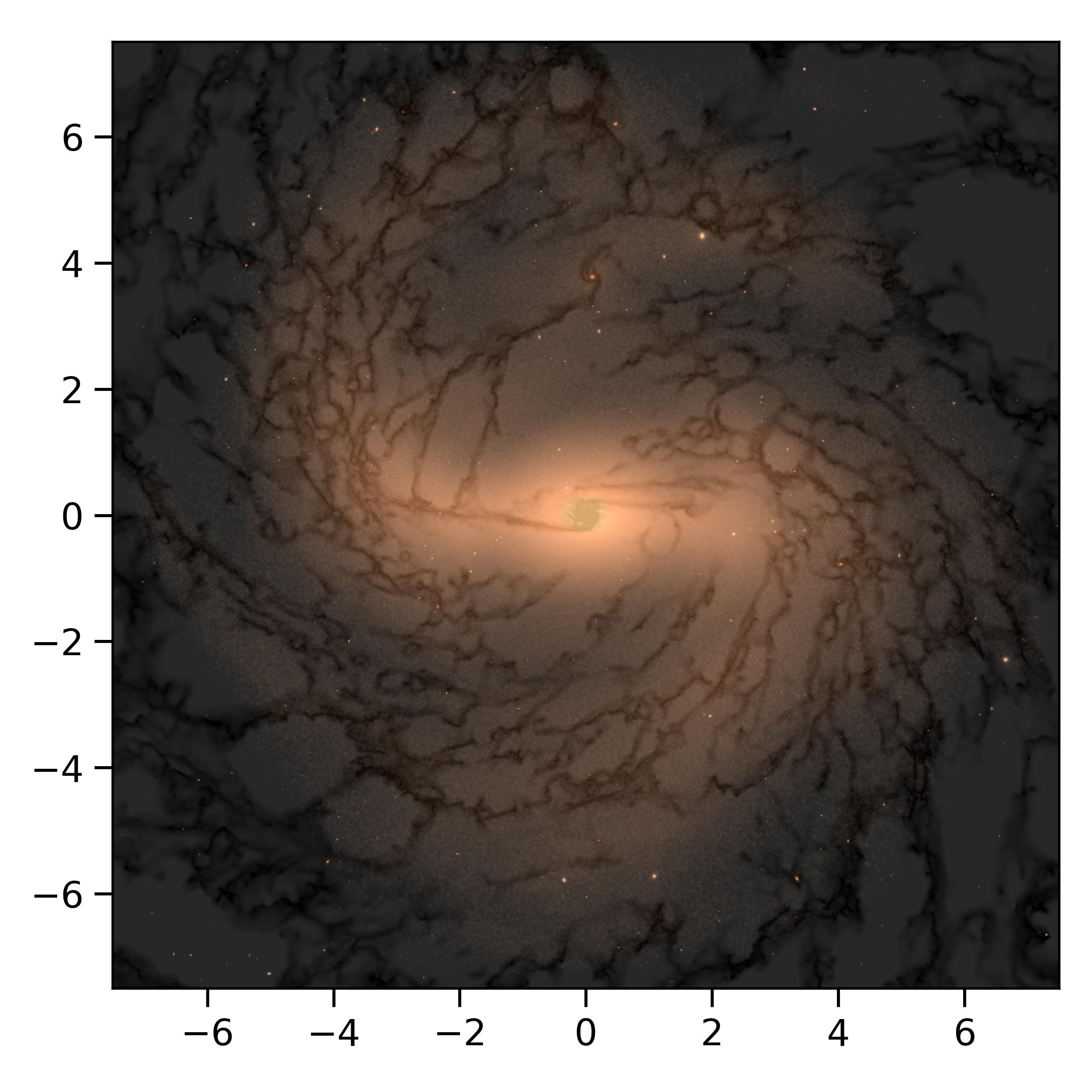}}\hspace{0pt} &
    \parbox[c]{0.32\textwidth}{\centering\includegraphics[width=0.32\textwidth]{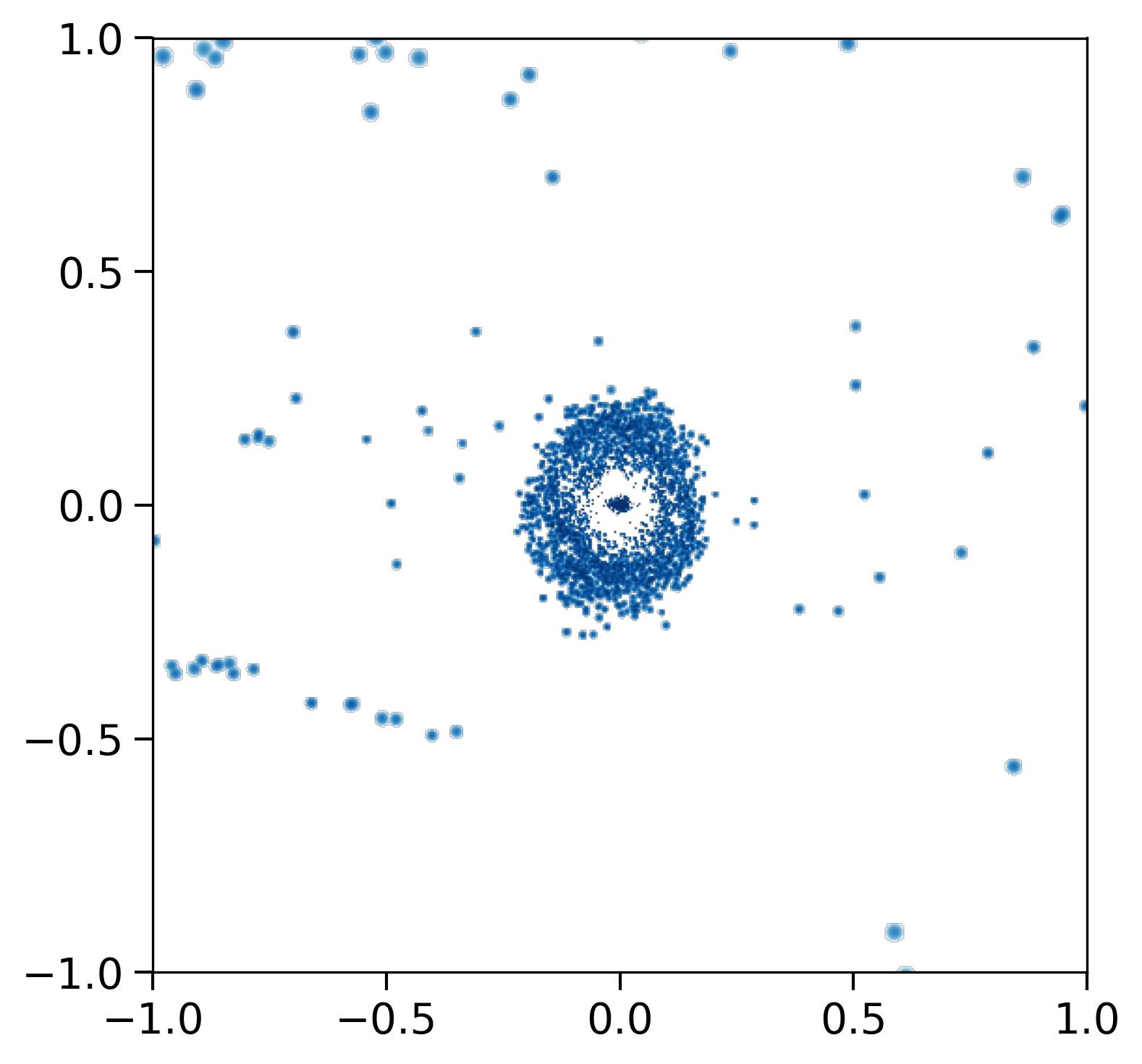}} \\[4pt]
 
    \multicolumn{3}{c}{\textbf{\large 2.6 Gyr}} \\[-1pt]
    \rotatebox{90}{\large \textbf{y [kpc]}}\hspace{-3pt} &
    \parbox[c]{0.34\textwidth}{\centering\includegraphics[width=0.34\textwidth]{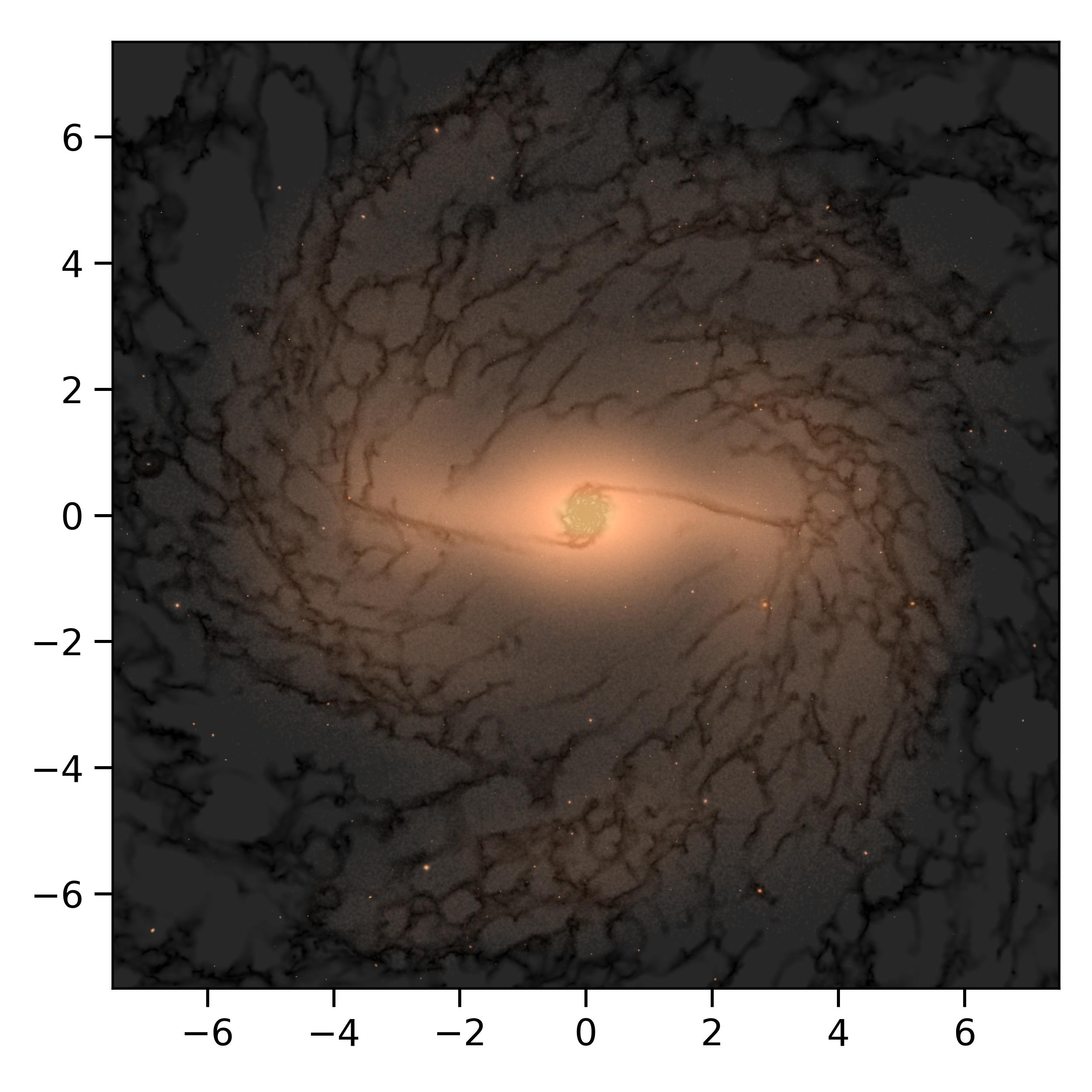}}\hspace{0pt} &
    \parbox[c]{0.32\textwidth}{\centering\includegraphics[width=0.32\textwidth]{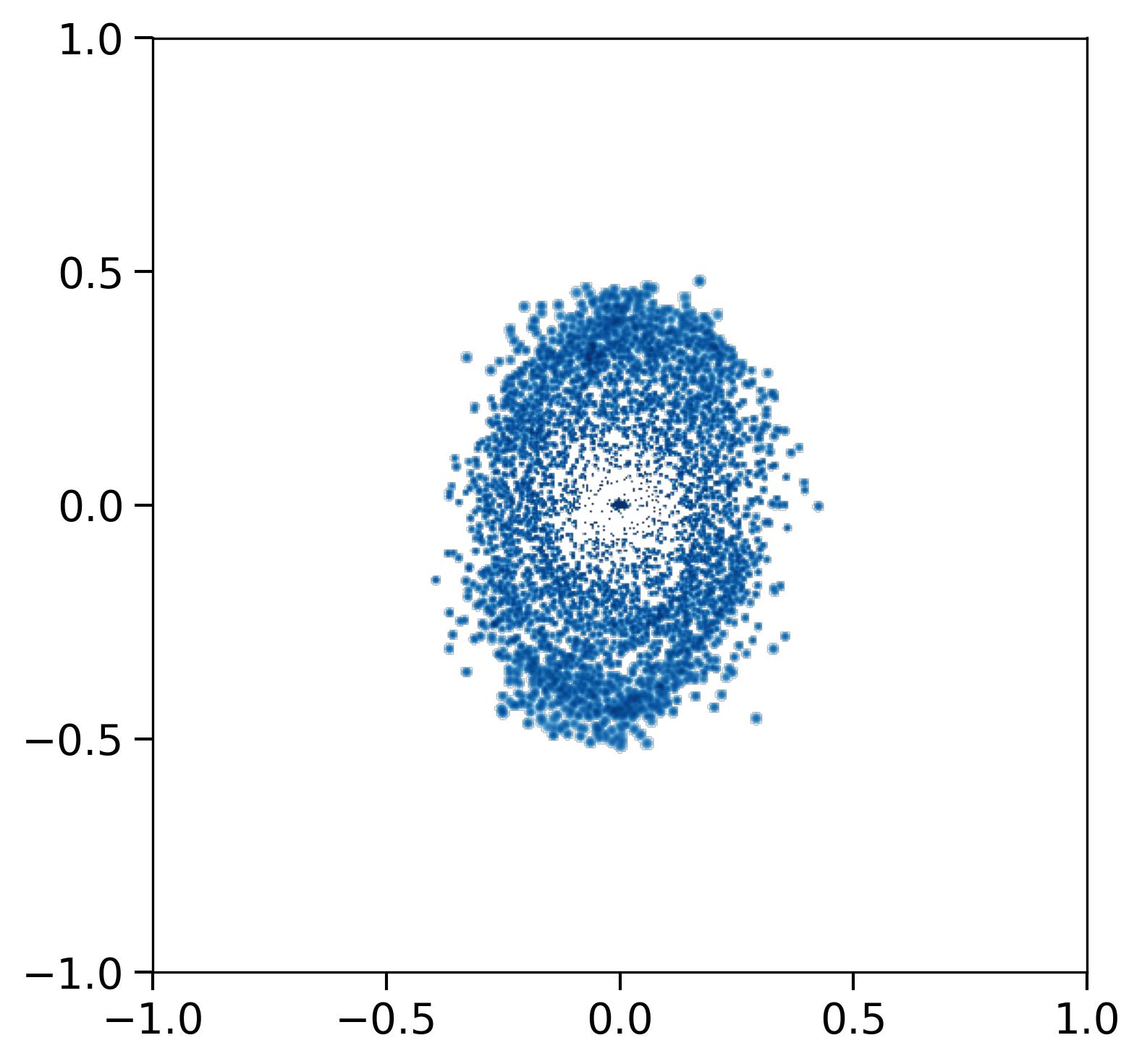}} \\[4pt]
 
    \multicolumn{3}{c}{\textbf{\large 4.0 Gyr}} \\[-1pt]
    \rotatebox{90}{\large \textbf{y [kpc]}}\hspace{-3pt} &
    \parbox[c]{0.34\textwidth}{\centering\includegraphics[width=0.34\textwidth]{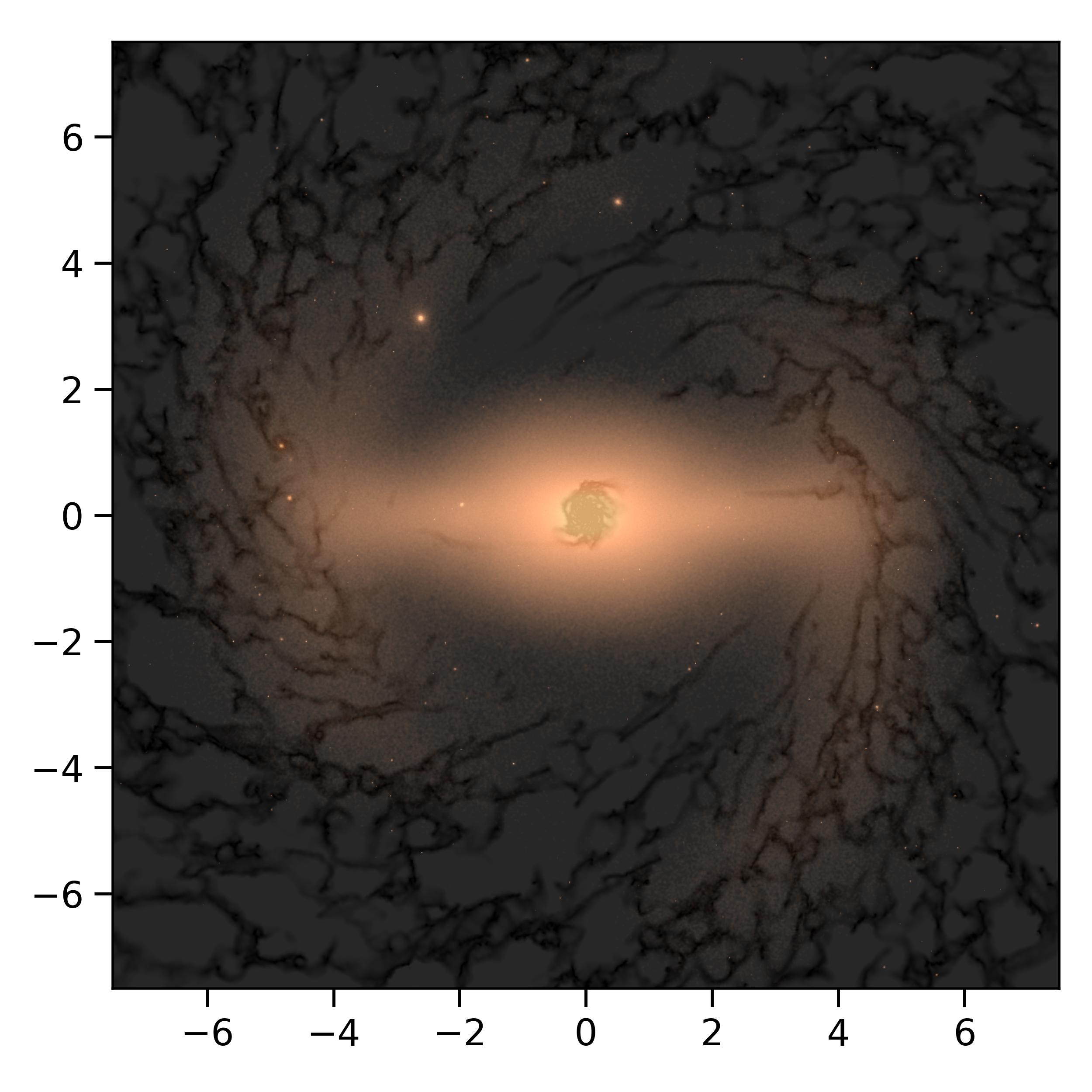}}\hspace{0pt} &
    \parbox[c]{0.32\textwidth}{\centering\includegraphics[width=0.32\textwidth]{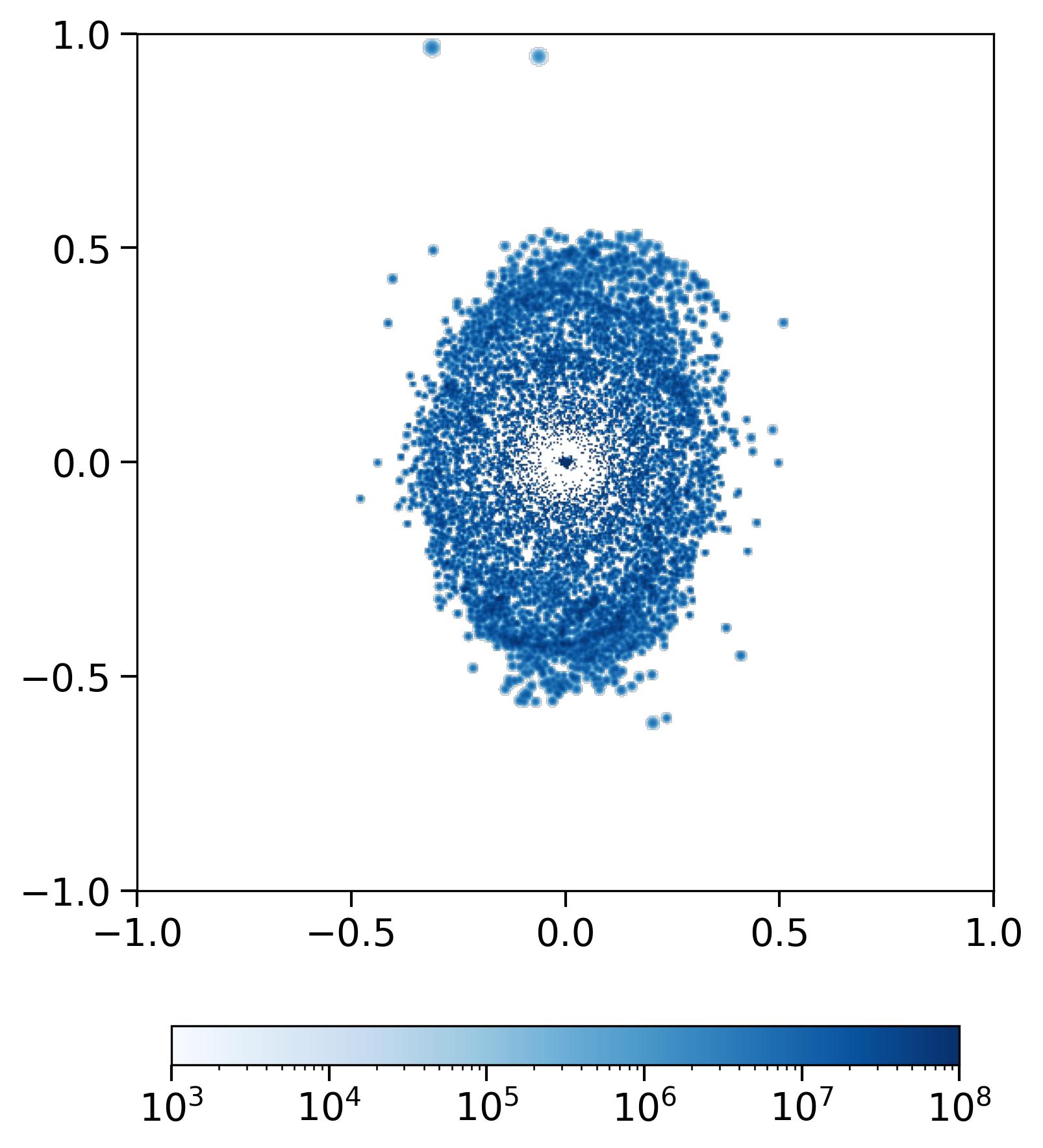}} \\[4pt]

    & \multicolumn{2}{c}{\textbf{\large x [kpc]}}
    \end{tabular}
    \caption{Left: Face-on projections of the stacked surface density distribution of stars and gas in the X-Y plane within a $15\times15 \ \kpc^2$ box at 1.5, 2.6, and 4.0 Gyr (from top to bottom). Right: Face-on projections of the surface density of blue stars (age $<0.1$ Gyr) in the X-Y plane within a $1\times1 \kpc^2$ box at 1.5, 2.6, and 4.0 Gyr. The color bar is in units of $\rm M_{\odot}\,kpc^{-2}$.}
    \label{fig:faceon}
\end{figure*}

In the first paper of the SMUGGLE-Ring project \citep[][hereafter Paper~I]{kwak26b}, we showed that varying the bulge-to-disk mass ratio ($0 < B/D < 0.091$) significantly affects the formation epoch, size, star formation rate, and kinematics of nuclear structures. These quantities fall within the range spanned by recent observations \citep{gadotti19, desafreitas23b, gadotti25, gleis26, leconte26} and have led to a proposed young-bar versus old-bar scenario for the Milky Way (MW), depending on bulge mass and bar age. As demonstrated in Paper~I and references therein, nuclear stellar disks (NSDs) are rotationally supported stellar systems that expand outward because bar-driven gas primarily forms stars at the outer edge of the nuclear gas disk, with that star-forming edge itself ending up migrating outward over time. 

In contrast, nuclear star clusters (NSCs) are dense and pressure-supported structures at the centers of galaxies \citep{neumayer20}. The formation mechanism of NSCs remains an active area of research, with two primary channels being (i) the inspiral of star clusters  (see e.g.  \citealt{hartmann2011}, \citealt{guillard16}, \citealt{Tsatsi2017}, \citealt{Abbate2018},  \citealt{Mastrobuono2021}) and (ii) in situ star formation driven by gas inflows  (see e.g. \citealt{fahrion22, fahrion24}, \citealt{Aharon2015}, \citealt{Brown2018}). For example, \cite{nogueras23} proposed that the nuclear structures in the MW may have originated from a single progenitor, yet the scaling relations between NSCs and NSDs presented by \cite{gadotti25} show no evidence of a shared formation history. In this paper, we present the first self-consistent simulation to investigate the formation of NSDs and NSCs and their evolutionary links in light of the accretion history \citep{fahrion22} and the stellar feedback-driven shocks \citep{leaman19,kolcu23,kolcu26}. The high resolution of our model allows us to trace in detail the growth and star formation history (SFH) of the nuclear structures.

\begin{figure*}[!t]
    \centering    \includegraphics[width=0.40\textwidth]{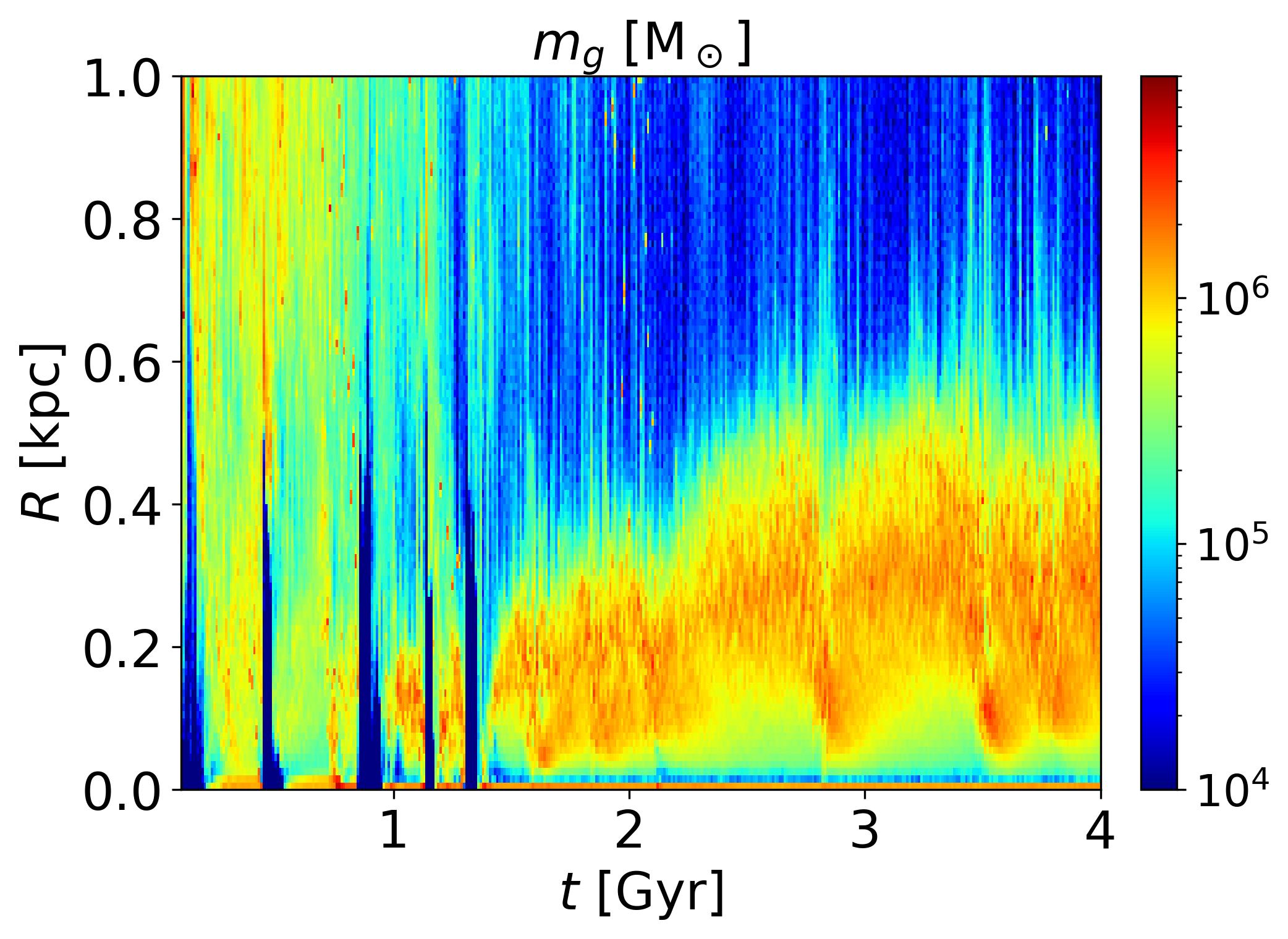}\includegraphics[width=0.40\textwidth]{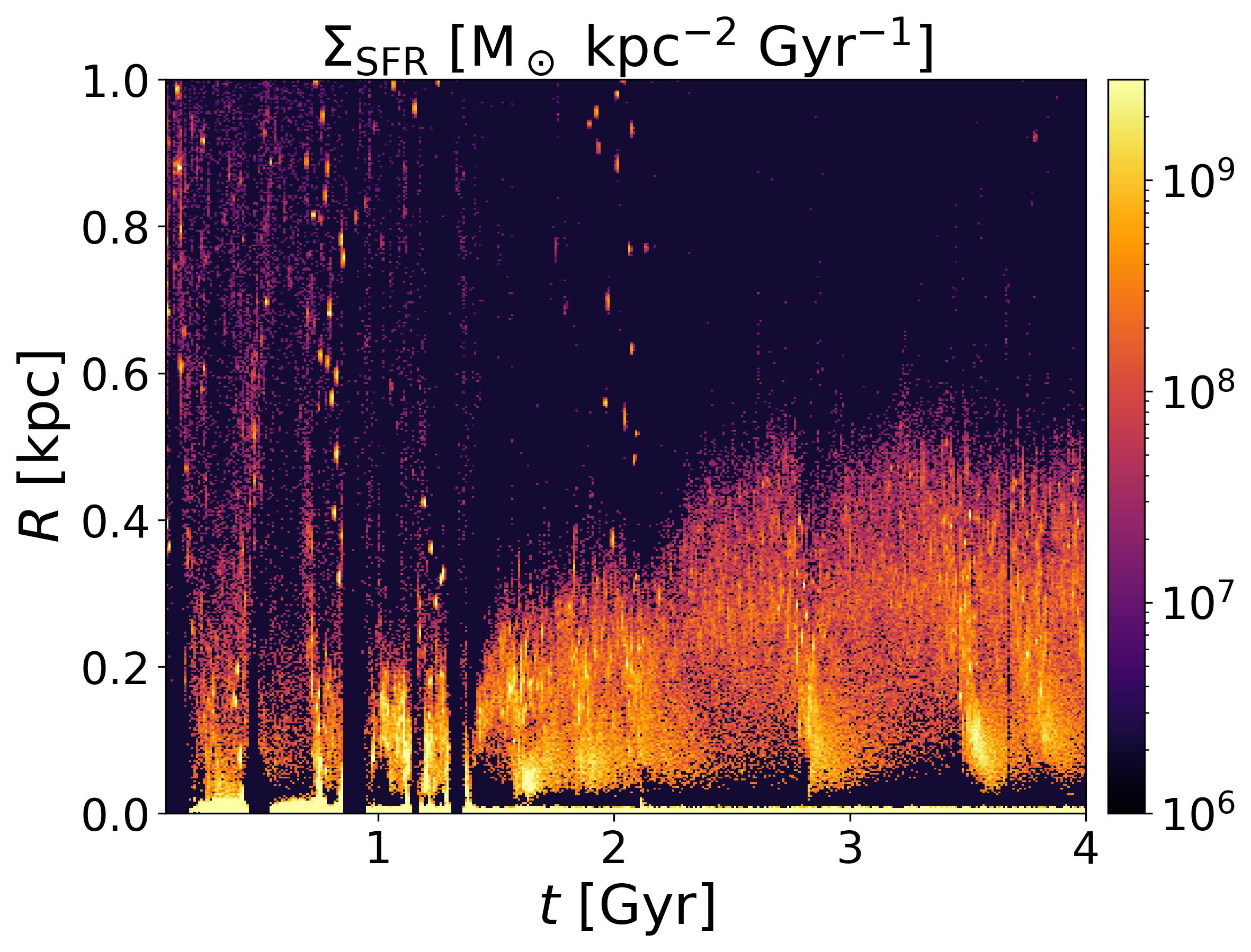}
    
    \caption{Left: Temporal evolution of radial distribution of gas mass, $m_g$, in each bin within 1 kpc for 4 Gyr. The bin size is 0.01 kpc. Right: Temporal evolution of star formation rate surface density $\Sigma_{\rm SFR}$ in each bin within 1 kpc. Multiple stellar feedback events drive gas inward and spikes star formation.}
    \label{fig:gas_sfr}
\end{figure*}

\section{Results}

\subsection{Isolated galaxy model}

For our galaxy model r2c14b05, we increased the resolution of model r1c14b05 from Paper I by a factor of 10 and evolved the system in isolation for 4 Gyr using the (Stars and MUltiphase Gas in GaLaxiEs (SMUGGLE) interstellar medium (ISM) and stellar feedback framework \citep{marinacci19}. Model r1c14b05 is a MW-mass galaxy consisting of a stellar disk, a gas disk, a stellar bulge, and a DM halo. The baryonic and DM particles have masses of $10^{3} \Msun$ and $10^{4} \Msun$, respectively. We refer to the stellar particles in the initial conditions as initial stars, to the newly formed stars as new stars, and to the stars younger than 0.1 Gyr as blue stars. More details are provided in Appendix~\ref{appendix:ic} and Paper~I.

\subsection{Bar formation}
In line with the results of Paper~I, our galaxy model naturally develops a bar around 1 Gyr without imposing a fixed bar potential. This enabled us to examine the time evolution of the bar and the resulting changes in the nuclear structures. By the end of the evolution, the bar grows stronger and larger, reaching around 5 kpc, a length comparable to that of the MW bar (e.g., \citealt{portail17}). It drives gas inflows toward the center, as shown in Fig. \ref{fig:map8kpc}, with details in Appendix \ref{appendix:fourier}.

\subsection{Nuclear structures}
The morphological evolution of the bar and nuclear structures is illustrated in Fig.~\ref{fig:faceon}. The left panel shows the face-on projections of the stacked surface density distribution of all stars and gas within a $15 \times 15$ kpc$  ^2  $ box at 1.5, 2.6, and 4.0 Gyr (see Fig.~A.3 in Paper~I). As the bar rotates and dynamically interacts with the DM halo, it becomes more prominent and elongated by driving disk stars onto bar-supporting orbits, thereby leaving low-density zones (dark gaps) in the inner disk \citep{athanassoula03,kwak17,kim25}. At the corresponding times, the face-on projections of surface density of blue stars in the bottom panel of Fig.~\ref{fig:faceon} shows the formation of the star forming nuclear ring-and-disk, which is relatively small and nearly circular at 1.5 Gyr, but its ellipticity briefly increases to $e\approx0.7$ (also shown in Paper~I). However, owing to the enhanced bar length and strength in our high-resolution model, the bar persistently brings more gas from the outer disk region into the inner kiloparsec (Fig.~\ref{fig:map8kpc}). This delays gas depletion in the nuclear gas disk, allowing the NSD to maintain ongoing star formation activity, instead of forming a nuclear stellar ring.

Figure~\ref{fig:gas_sfr} shows the time evolution of the radial distribution of the projected gas mass $m_g(r)$ within 1 kpc (radial bin size of 0.01 kpc). The nuclear gas disk begins to form shortly after the bar formation at around $  t_{\rm bar}=1  $ Gyr and stabilizes around 1.5 Gyr after several disruptions from stellar feedback processes. Although the nuclear gas disk grows in size, stellar feedback repeatedly drives gas from its outer edge toward the center. For instance, cavities appear in the outer part of the gas distribution at 1.6, 2.8, and 3.5 Gyr. At these times, the inner region of the gas disk ($R<0.1$ kpc) shows a clear increase in mass. This repeated feedback effect redistributes the angular momentum of the nuclear gas disk and drives gas toward the inner region, so the distribution of blue stars exhibits a more disk-like morphology at 2.6 and 4.0 Gyr in Fig.~\ref{fig:faceon}. These feedback-driven shocks generate rapid increases in the star formation rate surface density $  \Sigma_{\rm SFR}  $ (Fig.~\ref{fig:gas_sfr}). Interestingly, \cite{kolcu23} found that NGC 1097 is indeed experiencing repeated shocks that drive gas inflows into the nuclear region. Observationally, such shock signatures are common in barred galaxies \citep{kolcu26}, and we present one of the first detailed numerical simulation on the effects of this process for the growth of nuclear structures.

\begin{figure*}[!t]
    \centering
    
    \includegraphics[width=0.40\textwidth]{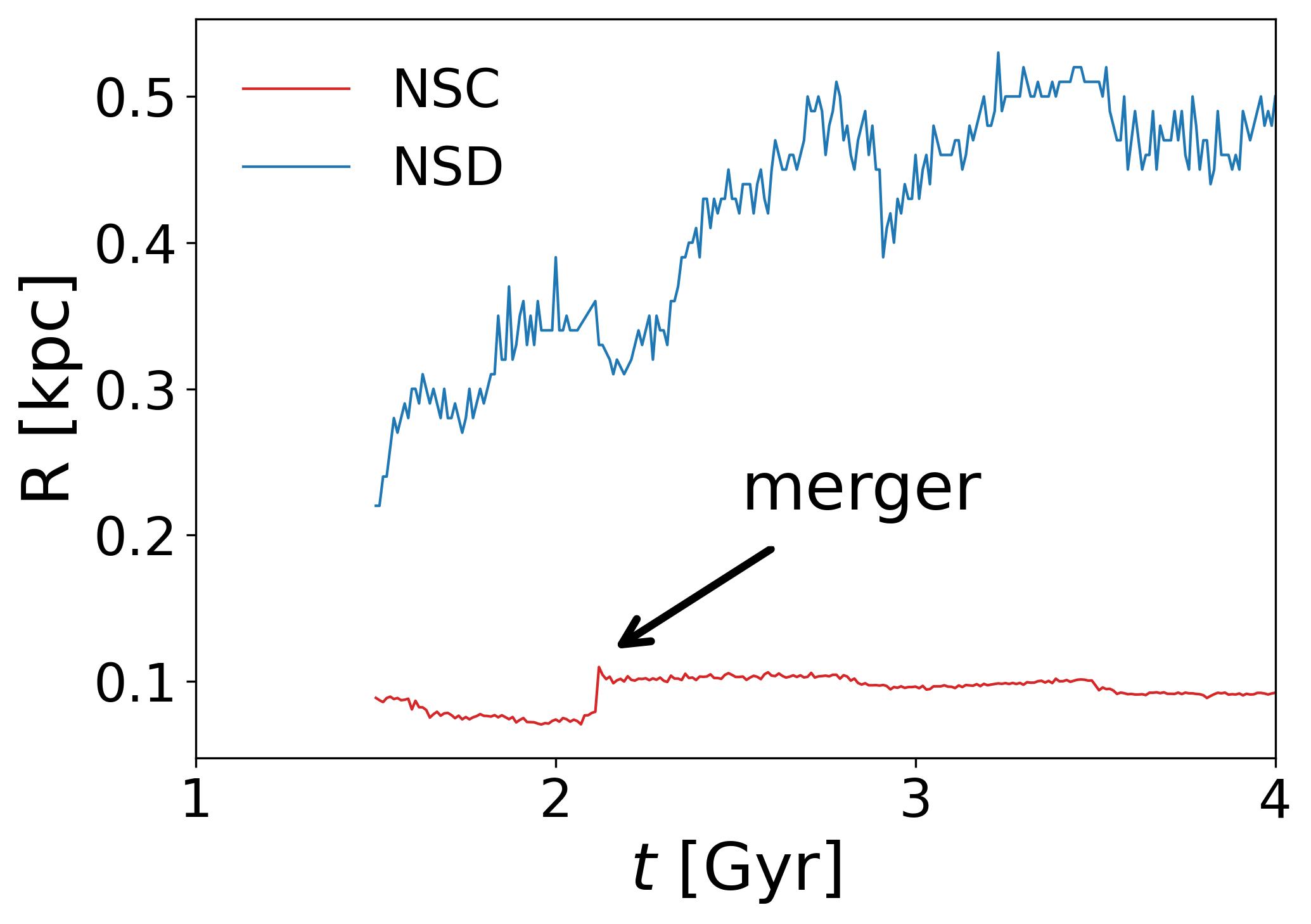}\includegraphics[width=0.40\textwidth]{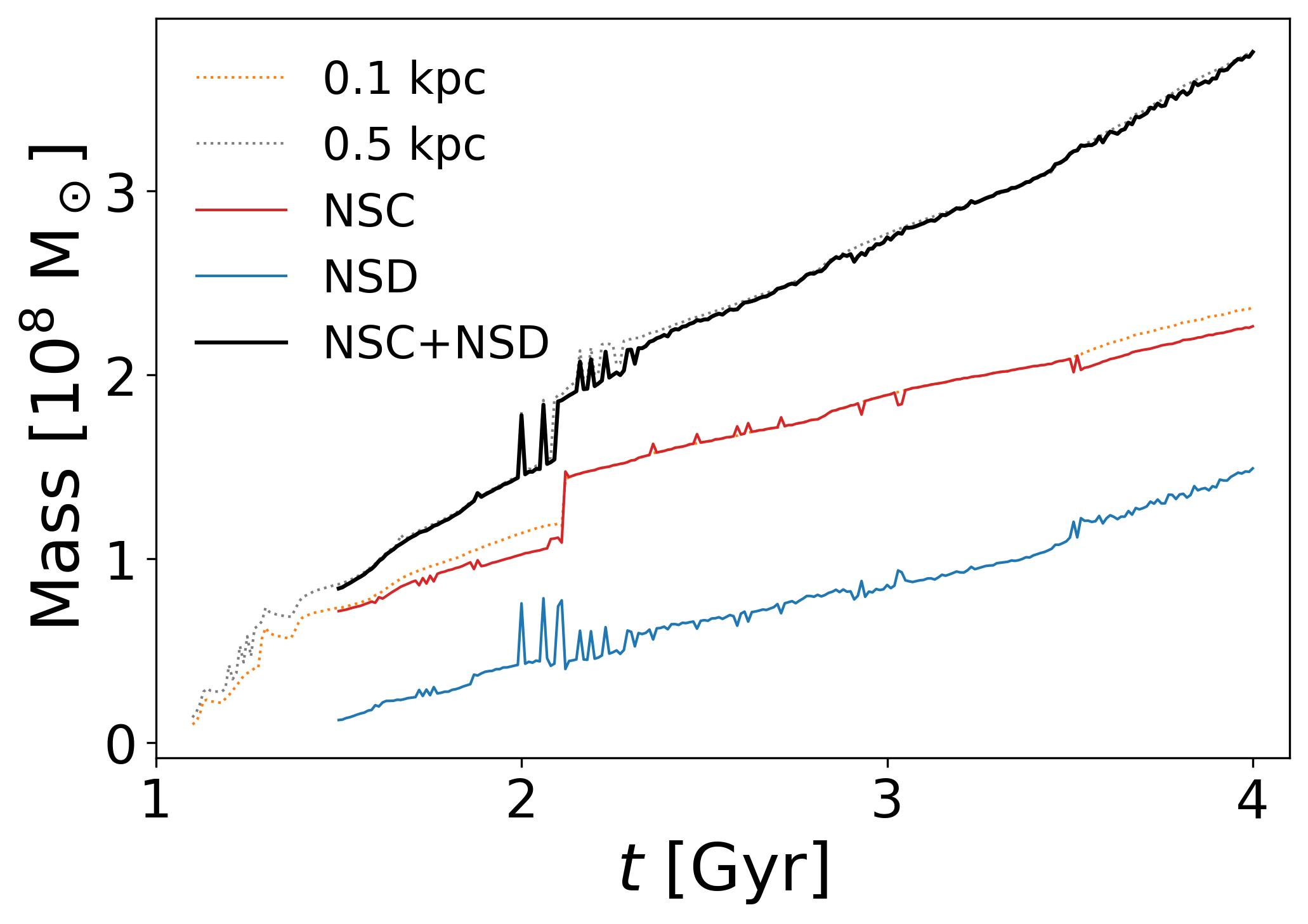}

    \includegraphics[width=0.40\textwidth]{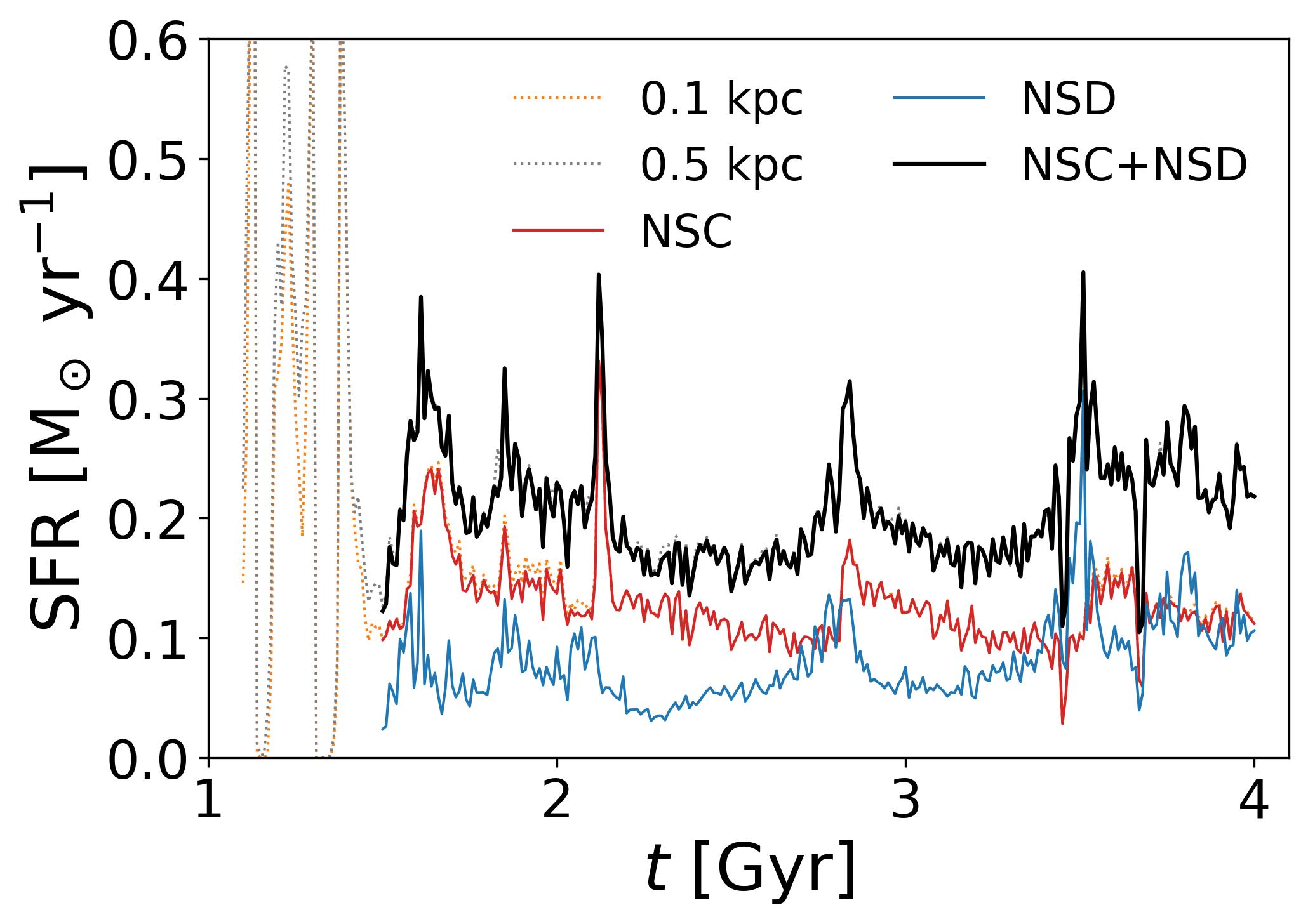}\includegraphics[width=0.40\textwidth]{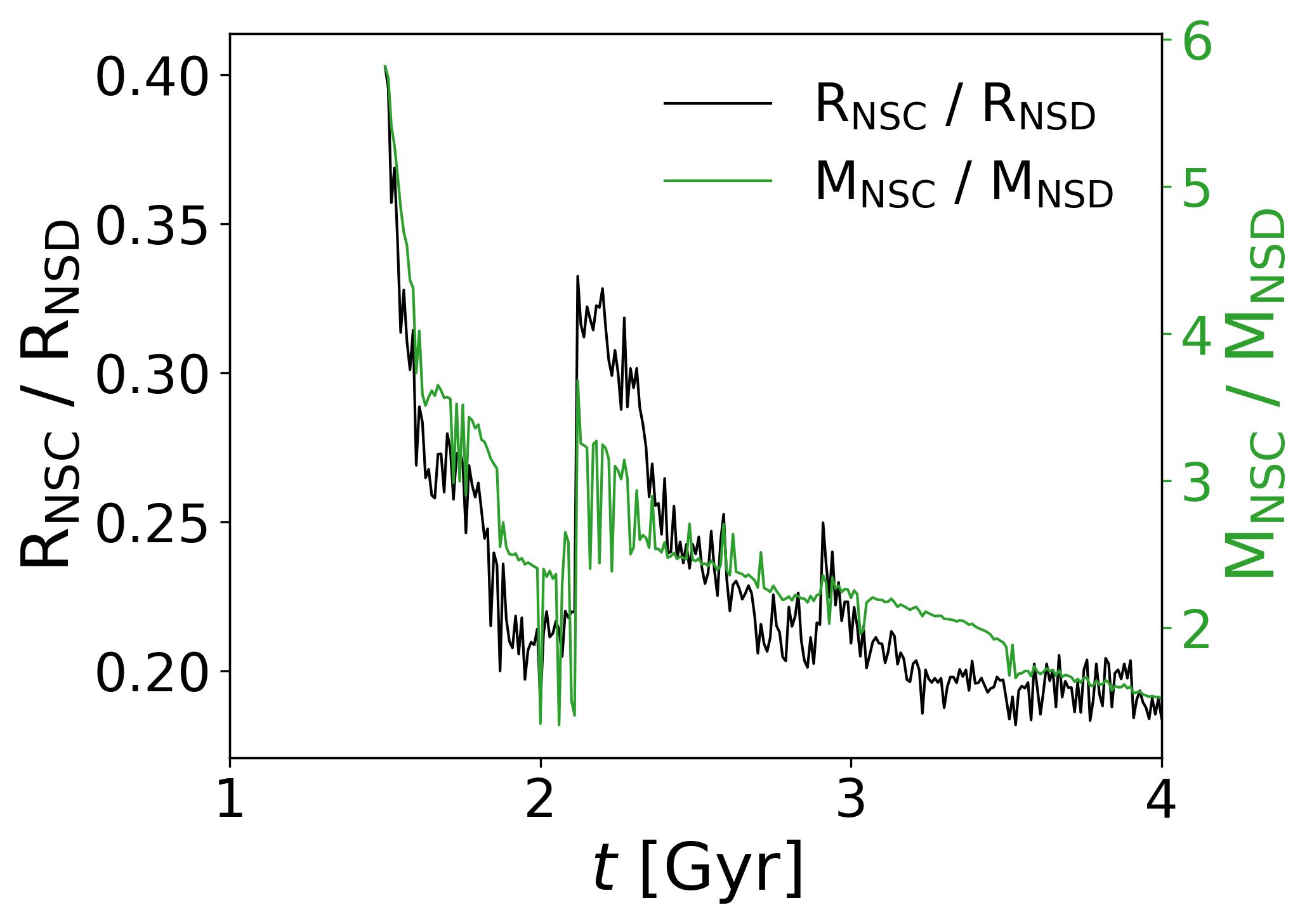}

    \caption{Temporal evolution of properties of nuclear structures for 4 Gyr. Top-left: Radius of the NSC and NSD. The pressure-supported region with $  v_{\phi}/\sigma < 0.3  $ is defined as the NSC. Top-right: Mass of the NSC, NSD, and their sum. Masses within 0.1 and 0.5 kpc are overlaid from 1.1 Gyr for comparison. The new stars born after $t_{\rm bar}=1\Gyr$ are considered for the mass and $v_{\phi}/\sigma$ calculation. Bottom-left: SFRs of the NSC, NSD, and their sum. The star formation rates within 0.1 and 0.5 kpc are overlaid from 1.1 Gyr, shortly after bar formation but before the emergence of stable nuclear structures. Bottom-right: Radius and mass ratios of the NSC and NSD.}
    \label{fig:nscnsd}
\end{figure*}

These repeated feedback-driven shocks also regulate the size of the NSD. The top-left panel of Fig.~\ref{fig:nscnsd} illustrates the time evolution of the NSD and NSC radii. We define the NSD radius as the outermost annulus (bin width of 0.01 kpc) that contains 50 blue stars (age $<0.1\Gyr$). For example, when the feedback event is present at 2.8 Gyr, the NSD radius temporarily decreases from 0.5 to 0.4 kpc.

We dynamically defined the NSC as the pressure-supported nuclear region where $v_{\phi}/\sigma<0.3$\footnote{Using the threshold $  v_{\phi}/\sigma < 0.5  $ increases the NSC size by 0.03 kpc.}. This is in contrast to Paper~I, where the NSC was defined as the region inside the semiminor axis of the nuclear stellar ring. Over time, the NSC region remained within 0.1 kpc except a sudden increase at 2.1 Gyr by a merger of massive star cluster. We note that only new stars born after bar formation ($  t_{\rm bar}=1  $ Gyr) were considered when calculating the mass and kinematics of the nuclear structures. In Paper~I, the radial profiles of the entire new stars still exhibited clear kinematic differences from those of the initial stars. In this work, we additionally excluded new stars born before $t_{\rm bar}=1$ Gyr, following the age-dating method for bars using nuclear disks \citep{baba20, sanders24, desafreitas23a}. Figure~\ref{fig:nsd_nocut} shows the face-on projections of $  v_\phi  $ and $  v_\phi/\sigma  $ at 1.5, 2.6, and 4.0 Gyr without the age selection. In Fig.~\ref{fig:nsd_agecut}, we see that applying the age selection visibly reduces contamination from new stars unrelated to the nuclear structures, allowing even the small NSD and NSC at 1.5 Gyr to be clearly distinguished. Additionally, the youngest stellar population resides at the outer edge of the NSD (SFT profile in Fig.~\ref{fig:nsd_agecut}, where SFT stands for star formation time and denotes the birth time of each star particle in the simulation). The radial profiles of SFT, $v_\phi$, and $v_\phi/\sigma$ from 1.5 to 4.0 Gyr at 0.5 Gyr intervals are illustrated in Fig.~\ref{fig:profiles} after applying the age cut, revealing a clear inside-out trend in which the nuclear disk expands via star formation at its outer edges, causing the peak of the SFT to migrate outward over time \citep{bittner20,ryde25,schultheis25b}. During this inside-out formation, the metallicity gradients (Fig.~\ref{fig:profiles}) also naturally resemble the observed trends \citep{bittner20,schultheis25b}.

In the top-right and bottom-left panels of Fig.~\ref{fig:nscnsd}, we present the mass and star formation rate (SFR) of the NSC and NSD and compare them with measurements within fixed radii of 0.1 and 0.5 kpc. Both the size and SFR of the NSD are comparable to those measured in the PHANGS-ALMA survey ($R\sim0.4^{+0.25}_{-0.15}\,\kpc$ and $\rm SFR \sim0.21^{+0.15}_{-0.16}$ $\Msun$yr$^{-1}$ in \citealt{gleis26}). The temporal evolution of mass growth shows that the NSC and NSD generally co-evolve, except for the sudden increase in the NSC mass and radius around 2.1 Gyr caused by the merger of a massive star cluster. 
Bars are known to promote the formation of massive clusters \citep{ali23}. Using the friends of friends algorithm (e.g., \citealt{davis85}), we identified approximately 200 star clusters with masses above $10^5\Msun$ (see Appendix~\ref{appendix:starcluster}). The most massive of these has a mass of approximately $3 \times 10^{7} \Msun$ and becomes trapped by the bar. As its orbit decays via dynamical friction, it enters the NSD, orbits within the NSD several times, and merges with the NSC at 2.1 Gyr (Fig.~\ref{fig:accretion}). Consequently, both the mass ratio $M_{\rm NSC}/M_{\rm NSD}$ and the radius ratio $R_{\rm NSC}/R_{\rm NSD}$ increase abruptly as a result of this merger event (Fig.~\ref{fig:nscnsd}). Such cluster inspirals are important for reproducing the kinematical properties of the MW's NSC \citep{tsatsi17}.

The SFR trend is also similar between the NSC and NSD, exhibiting spikes when stellar feedback events drive shocks on the outer region of the nuclear gas disk (Fig.~\ref{fig:nscnsd}). These spikes appear with a slight delay in the NSC relative to those in the NSD. The merger of the massive star cluster spikes the SFR only in the NSC. Overall, the masses and SFRs of the NSD and NSC co-evolve unless a merger event occurs. In environments with sustained external gas inflow and/or gas-rich galaxy mergers, more massive star clusters may form \citep{li22}, which can contaminate the intrinsic link between NSCs and NSDs.

\section{Discussion and summary}
This work presents the first self-consistent hydrodynamical simulation that examines the detailed formation and evolution of the NSD and NSC in a MW-mass galaxy. Most of the literature on nuclear structures is based on local simulations (e.g., \citealt{moon23}) that employ a fixed bar potential, whereas our system naturally forms a realistic bar that evolves over time through dynamical interactions with the gas and DM halo, thereby driving gas inflows from the disk regions. With realistic multiphase ISM and stellar feedback, our simulation naturally reproduces observed features in the nuclear regions of barred galaxies (e.g., gas cavities by the feedback in NGC 3351; \citealt{leaman19}). The feedback events generate shocks and redistribute angular momentum, thereby enhancing star formation in the NSC \citep{kolcu23, kolcu26}. In the local simulation of nuclear ring \citep{moon23}, such mass inflow and star formation in the NSC region do not occur unless a strong magnetic field is present. Aligning with Paper~I, our results support the inside-out formation scenario for nuclear structures as the NSD expands outward over time with its star-forming edge (e.g., SFT in Fig.~\ref{fig:nsd_agecut}). Although the NSC grows mostly via in situ star formation, its mass and size are affected by the merger with a massive star cluster.

Based on the NSD mass-size relation and the relation between NSD mass and galaxy mass, \cite{gadotti25} suggested that the formation of NSDs and NSCs is not directly connected because the corresponding scaling relations for NSCs do not agree with those for NSDs. Our results indicate that the observed mass scaling relations of NSDs and NSCs can be affected by the evolutionary timescale of nuclear structures as well as by the accretion history of star clusters. In secular evolution, the longer the evolution time the greater divergence in the ratio of mass and size between the NSC and the NSD. For example, the ratio $R_{\rm NSC}/R_{\rm NSD}$ decreases from 0.4 to 0.2 concurrently with the ratio $M_{\rm NSC}/M_{\rm NSD}$ decreasing from 6 to 2 (Fig.~\ref{fig:nscnsd}), while the galaxy mass remains the same. Thus, considering the time factor such as the age of the bar, which is directly related to the age and size of the NSD, could result in tighter scaling relations.

Our results suggest that the MW may host a relatively low-mass classical bulge. In our model with $B/D \approx 0.045$, the smallest radius of the NSD is 0.21 kpc, whereas the MW's NSD is significantly more compact, with a radius of $\sim 0.1$ kpc (\citealt{schultheis25a}, and references therein). Furthermore, the NSD in our simulation grows to $\sim0.5$ kpc around 3 Gyr, which is 2 Gyr after bar formation. In line with Paper I, we suggest that to explain the compact size of the MW's NSD, its spheroidal bulge would need to have a mass ratio lower than $B/D \approx 0.045$ \citep{nepal26}, favoring the assumption that the Galactic bar might be younger \citep{nepal24}. In future works, we will revisit our constraints on the bulge mass by including the effects of magnetic fields, which could alter the SFH and evolutionary timescale of NSDs.

Our NSC is also larger than the MW's NSC for the same reason. In the top panel of Fig. \ref{fig:accretion}, the NSC at the center appears resembles a massive star cluster, rather than a dispersed background stellar distribution. Our kinematical definition of the NSC ($v_{\phi}/\sigma<0.3$) aptly measures the compact region ($\sim$100 pc). Such a distinction around 100 pc is also seen in metallicity and kinematics (Fig.~\ref{fig:profiles}). For example, NGC 1433's NSD is $\sim$460 pc  \citep{combes13}, comparable to our NSD and its NSC size is measured as a Plummer scale radius of $35.8\pm8.9$ pc \citep{vermot23}, which gives $107.4\mathrm{pc}$ for approximately 90 percent of the total projected mass. We expect galaxies with larger NSDs and larger bulges to have more massive NSCs.

Finally, the mass of our nuclear structures ($M_{\rm NSC+NSD} = 0.9  $--$3.7 \times 10^{8}\Msun$ with $R_{\rm NSD}=0.21\sim0.51\kpc$) is notably lower than that of the MW estimated by \citet[$M_{\rm NSD} = 10.5^{+1.1}_{-1.0} \times 10^{8}\Msun$]{sormani22}. We note that the initial stars and blue stars (age $< t_{\rm bar}$) were not included in our mass calculation. Their combined mass within 0.5 kpc reaches $2.5 \times 10^{9}\Msun$. We suggest that applying selection criteria based on dynamics and age to the observed data would provide a more accurate estimate of the actual NSD mass.

\begin{acknowledgements}
We appreciate the referee's constructive comments. We gratefully acknowledge the computing time made available for the SMUGGLE-Ring project on the high-performance computer "Lise" at the NHR Center NHR@ZIB. This center is jointly supported by the Federal Ministry of Research, Technology, and Space and the state governments participating in the NHR (www.nhr-verein.de). The work of W-T.K.\ was supported by the grant of the National Research Foundation of Korea (RS-2025-00517264). IM acknowledges support by the Deutsche Forschungsgemeinschaft under the grant MI 2009/2-1. HL is supported by the National Key R\&D Program of China No. 2023YFB3002502, the National Natural Science Foundation of China under No. 12373006 and 12533004, and the China Manned Space Program with grant No. CMS-CSST-2025-A10.
\end{acknowledgements}

\bibliographystyle{aa}
\bibliography{ref}

\begin{appendix} 

\section{Method}\label{appendix:method}
\subsection{Initial condition}\label{appendix:ic}
The galaxy model r2c14b05 has the same density structure as the r1c14b05 model in \cite{kwak26b}, except that the number of particles is increased by a factor of 10. We construct an isolated galaxy model consisting of a stellar disk, a gas disk, a stellar bulge, and a DM halo. Our choice of structural parameters is comparable to those of a MW-mass galaxy. Before generating the gas disk, we first construct a collisionless system using the \textsc{galic} code \citep{yurin14}, which iteratively achieves an equilibrium configuration by adjusting the particle velocities. Afterward, a fraction of the stellar disk particles is converted into gas particles to form the gas disk in our initial conditions. As described in \cite{kwak26b}, we evolve the initial condition for $0.1\,\mathrm{Gyr}$ without star formation, allowing the galaxy model to relax into a quasi-equilibrium state.

We first construct a collisionless disk that follows the exponential density profile,
\begin{equation}\label{eq:expdisk}
\rho_{d} (R, z) = \frac{M_{d}}{4\pi z_{d} R_{d}^2} \exp \left( -\frac{R}{R_{d}} \right) \text{sech}^2 \left(\frac{z}{z_{d}}\right),
\end{equation}
where $R_{d}$ is the radial scale length, $M_{d}$ is the total disk mass, and $z_{d}$ is the vertical scale height.
The galaxy model has $M_{d}=6\times10^{10} \, \Msun$ and $R_d=3.0 \, \kpc$ with $R_d/z_d=10$. The velocity anisotropy parameter is set to $f_R = \sigma_{R}^2 / \sigma_{z}^2 = 1.4$ initially. 

The DM halo follows the \cite{hernquist90} profile,
\begin{equation}\label{eq:hernquist}
\rho_\text{DM} (r) = \frac{M_\text{DM}}{2\pi} \frac{a_h}{r(r+a_h)^3} ,
\end{equation}
where the total mass and scale length are $M_\text{DM}$ and $a_h$, respectively. The halo concentration parameter, $c$, is defined through
\begin{equation}\label{eq:cc}
a_h = \frac{r_{200}}{c} \left[2 \ln(1+c)-\frac{c}{1+c}\right]^{1/2},
\end{equation}
where $r_{200}$ is the virial radius \citep{springel05}. The total DM mass is set to $M_\text{DM} = 1.14 \times 10^{12} \, \Msun$ with $c=14$.

The stellar bulge follows the same Hernquist profile,
\begin{equation}\label{eq:hernquist}
\rho_b (r) = \frac{M_b}{2\pi} \frac{a_b}{r(r+a_b)^3} .
\end{equation}
The bulge mass is $M_b=2.5\times10^9 \, \Msun$. The bulge scale length is fixed at $a_b = 0.40\,\mathrm{kpc}$, which corresponds to a 3D half-mass radius of $\sim 0.97\,\mathrm{kpc}$ and a 2D projected effective radius of $\sim 0.73\,\mathrm{kpc}$.

For the collisional gas disk, a gas mass fraction $M_g/M_s \approx 9.1\%$ ($M_g=5 \times 10^{9} \, \Msun$) of the collisionless disk is converted into gas particles, where $M_g$ and $M_s$ are the masses of the gas and stellar disks, respectively. Consequently, the final galaxy model has a stellar disk mass of $M_s = 5.5 \times 10^{10} \, \Msun$, and the gas and stellar disks share the same density structure (see \citealt{kwak26a} for details). The gas disk is initialized with a temperature of $  10^{4}\,\mathrm{K}  $ and a uniform metallicity of $  Z = 0.3\,Z_\odot  $.

The galaxy model contains the following number of particles. stellar disk: $N_s = 5.5 \times 10^7$; gas disk: $N_g = 0.5 \times 10^7$; bulge: $N_b=2.5 \times 10^6$; and DM halo: $N_{\rm DM} = 11.4 \times 10^7$. To minimize gravitational heating from massive DM particles, the baryon-to-DM particle mass ratio is set to 10 \citep{kwak25, kwak26a}. The gravitational softening lengths are $0.005\,\mathrm{kpc}$ for stars and $0.01\,\mathrm{kpc}$ for DM particles. The minimum adaptive softening lengths is $0.005\,\mathrm{kpc}$ for gas cells. This galaxy model is evolved in isolation for $4\,\mathrm{Gyr}$ without external gas accretion.

\subsection{The SMUGGLE framework}\label{appendix:smuggle}
The SMUGGLE model \citep{marinacci19} is an explicit, comprehensive stellar feedback framework  and implemented in the moving-mesh hydrodynamics code AREPO \citep{springel10b,weinberger20}. It treats the ISM as a true multiphase medium and injects feedback energy and momentum locally from star particles without relying on decoupled wind particles or artificial pressurization. This approach overcomes the limitations of traditional sub-grid effective equation-of-state models, which tend to produce overly smooth ISM morphologies and unresolved disk structures. By self-consistently generating the hot, warm, and cold phases of the ISM, SMUGGLE accurately regulates star formation, drives realistic galactic outflows, and captures the multiphase structure of the gas, particularly its dense phase at $  n \gtrsim 1\,\mathrm{cm}^{-3}  $.

Key physical processes in SMUGGLE include supernova feedback with momentum injection distributed radially within a defined superbubble radius, radiative feedback from young massive stars (photoionization imposing a temperature floor in H II regions and radiation pressure with infrared optical depth), and stellar winds from OB and AGB stars. Cooling and heating processes are modeled in detail down to $  T \approx 10\,\mathrm{K}  $, incorporating primordial, metal-line, fine-structure, and molecular cooling together with cosmic-ray and photoelectric heating. Star formation follows a stochastic, probabilistic prescription: gas cells above a density threshold of $  n \approx 100\,\mathrm{cm}^{-3}  $ and satisfying a gravitational boundedness criterion are converted into stellar particles with efficiency $\epsilon=0.01$. This combination of physics enables converged results at various mass resolutions and has been successfully applied to a wide range of galaxy formation and evolution studies. More details of the SMUGGLE model are provided in \cite{marinacci19}.

\section{Supplementary figures}\label{appendix:supplementary}

\begin{figure}[t!]
    \centering
    \includegraphics[width=0.4\textwidth]{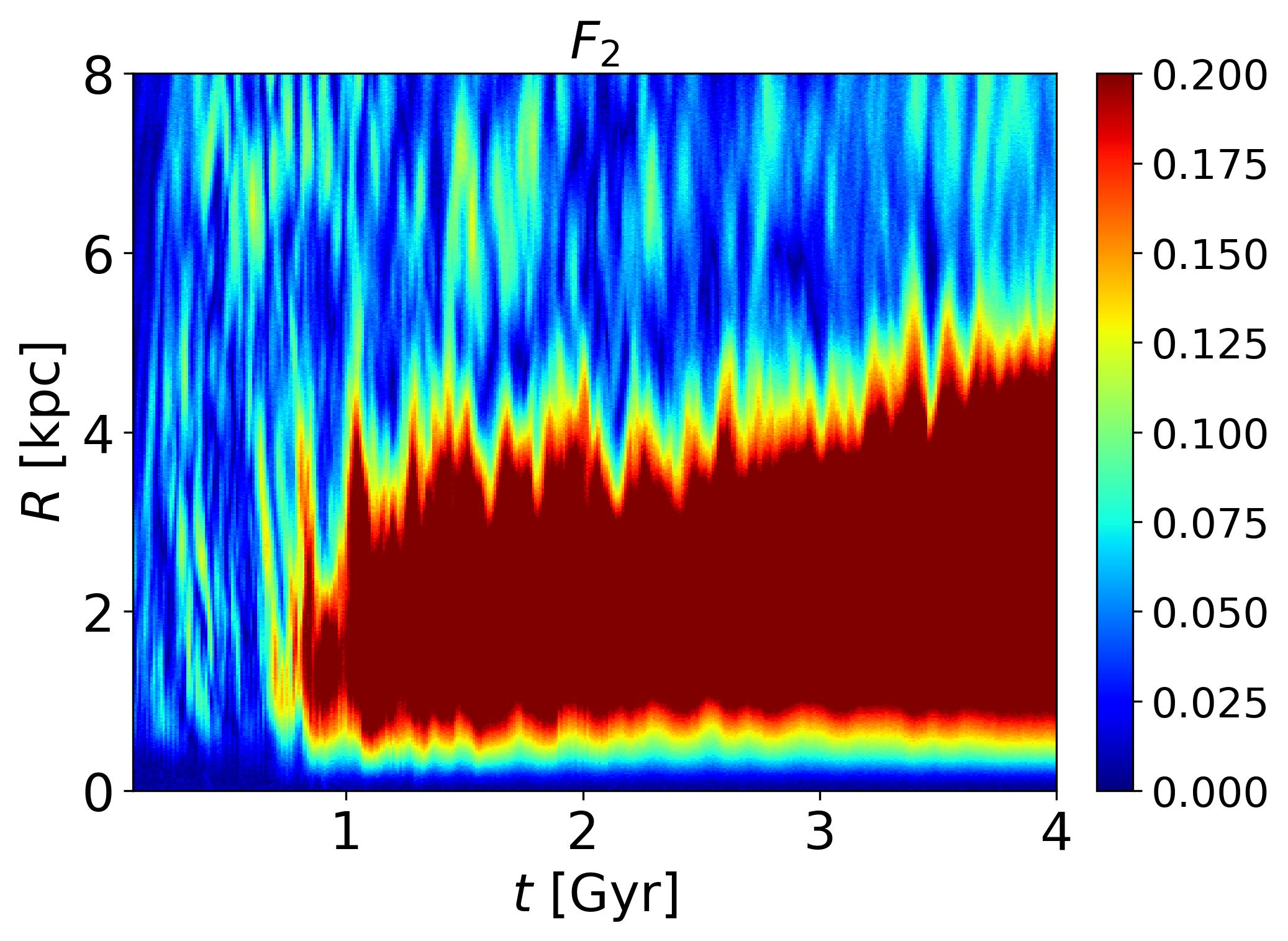}

    \includegraphics[width=0.4\textwidth]{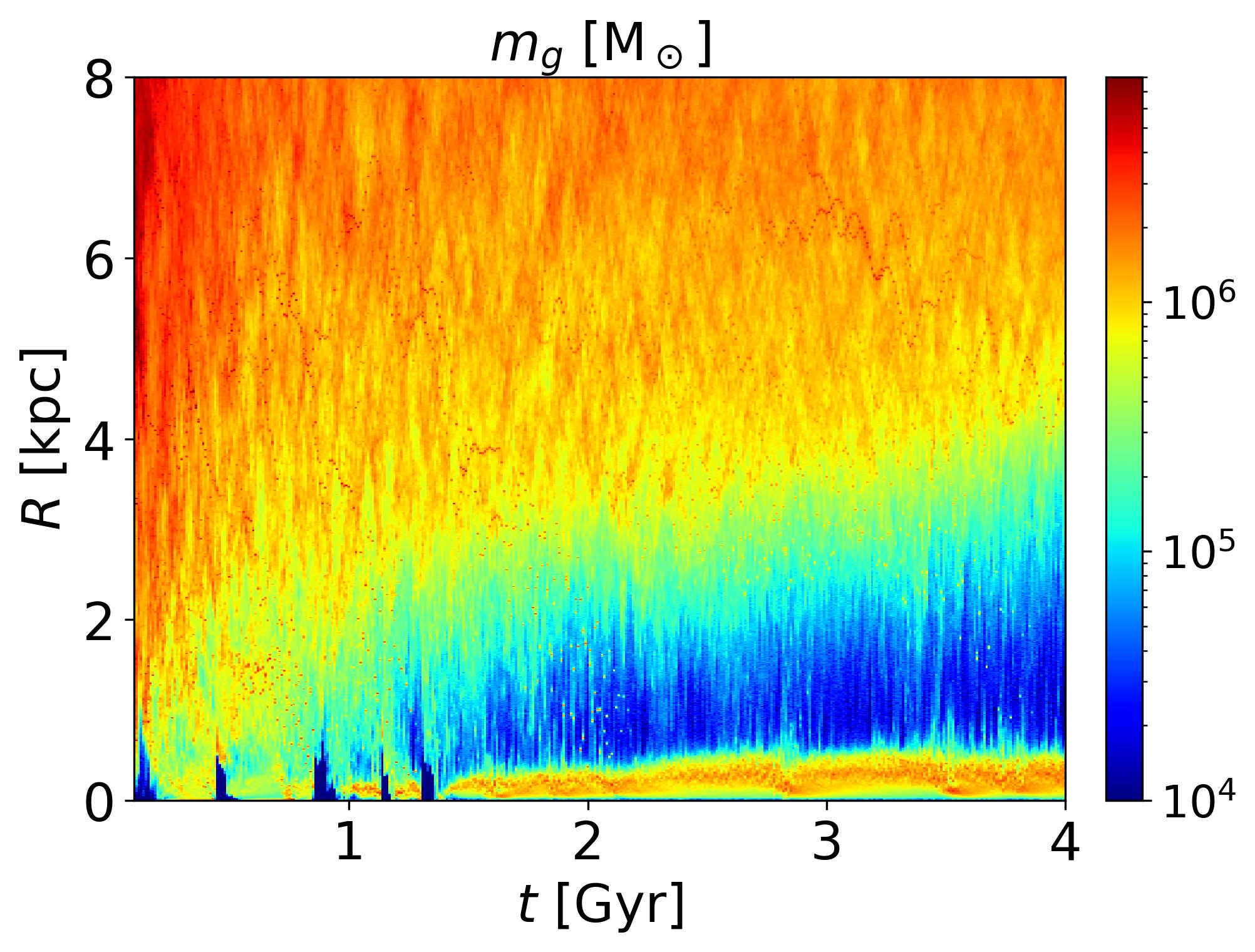}

    \caption{\textit{Top}: Temporal evolution of the radial distribution of $  F_2  $ within 8 kpc over 4 Gyr. The color bar is fixed from 0 to 0.2, with red indicating the bar region. \textit{Bottom}: Temporal evolution of the radial gas mass distribution within 8 kpc over 4 Gyr. The bin size for the radial profiles is 0.01 kpc.}
    \label{fig:map8kpc}
\end{figure}

\begin{figure}[t!]
    \centering
    \setlength{\tabcolsep}{0pt}
    \begin{tabular}{rcc}

    & \textbf{$v_{\phi}$ [km s$^{-1}$]} & \textbf{$v_{\phi}/\sigma$} \\[1pt]

    \rotatebox{90}{\quad \quad \quad 1.5 Gyr} &
    \includegraphics[width=0.22\textwidth]{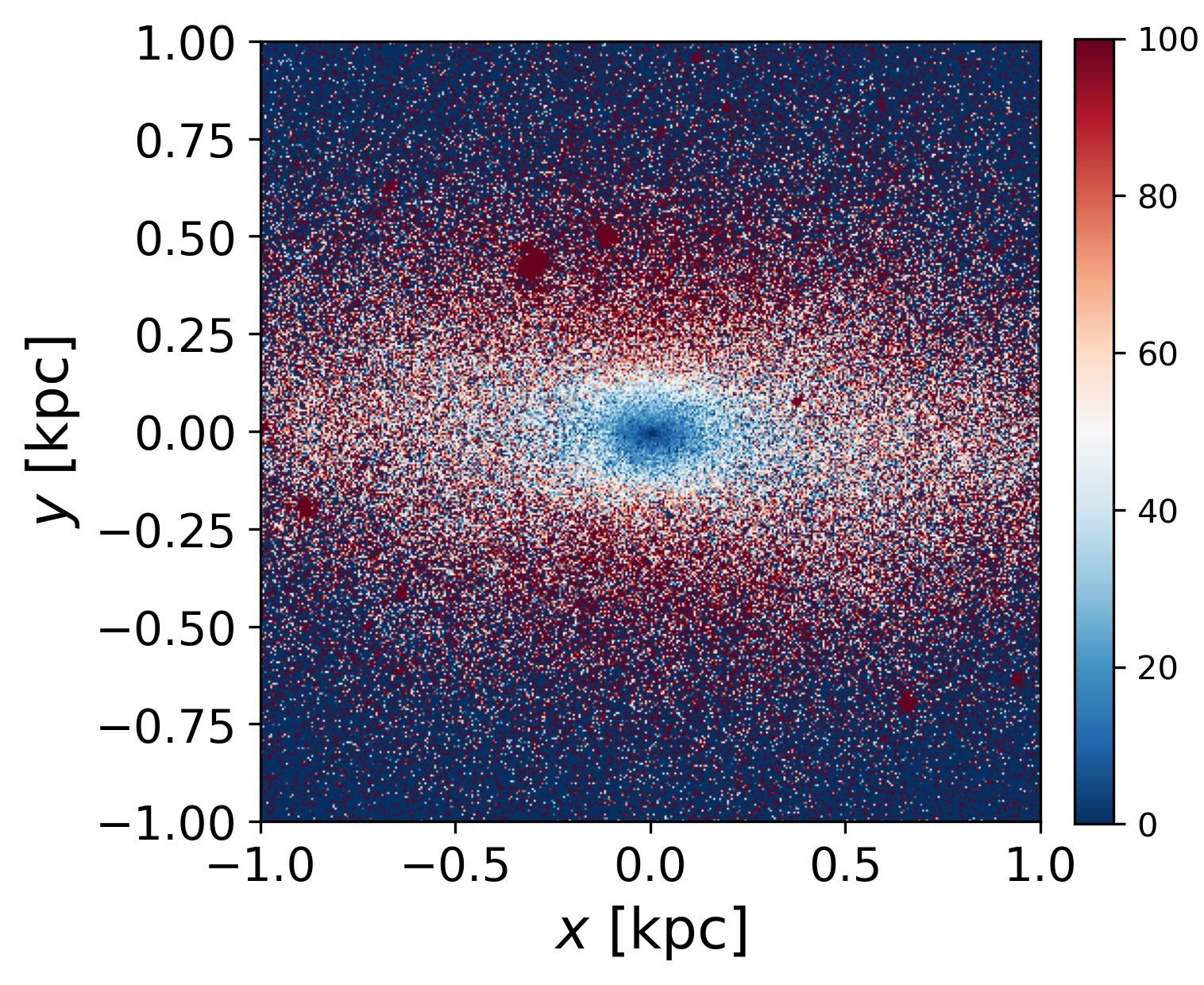} &
    \includegraphics[width=0.22\textwidth]{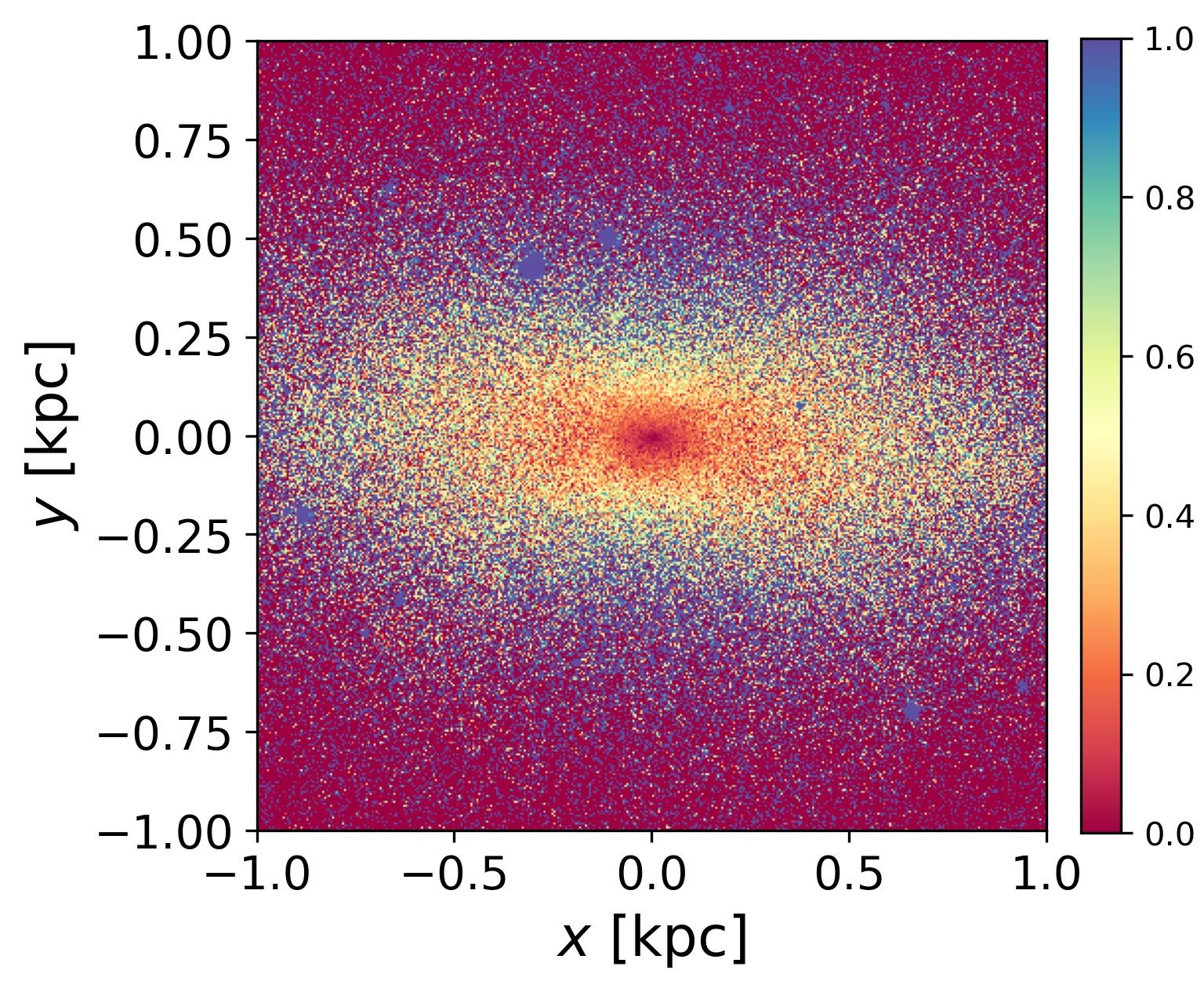} \\[14pt]

    \rotatebox{90}{\quad \quad \quad 2.6 Gyr} &
    \includegraphics[width=0.22\textwidth]{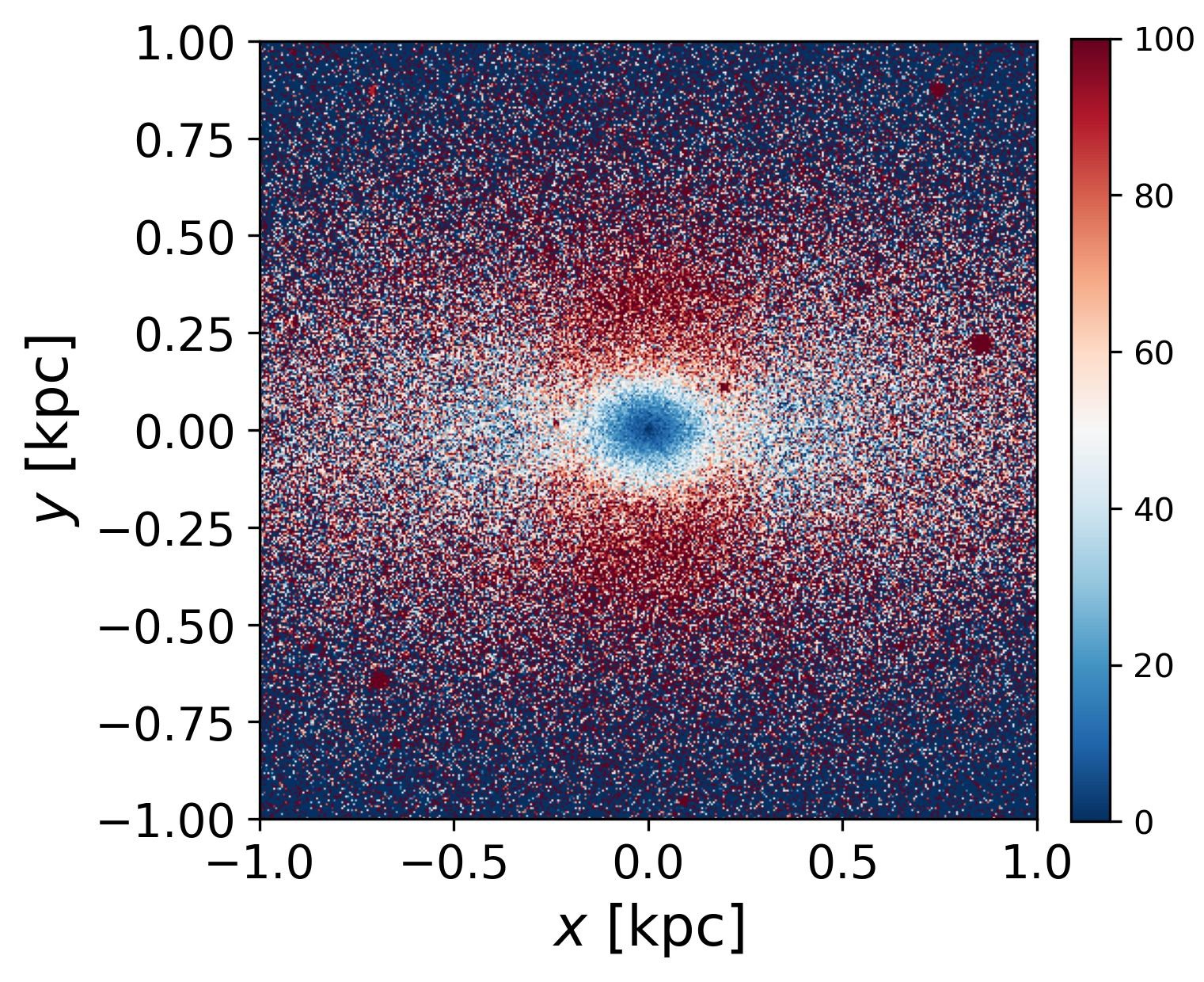} &
    \includegraphics[width=0.22\textwidth]{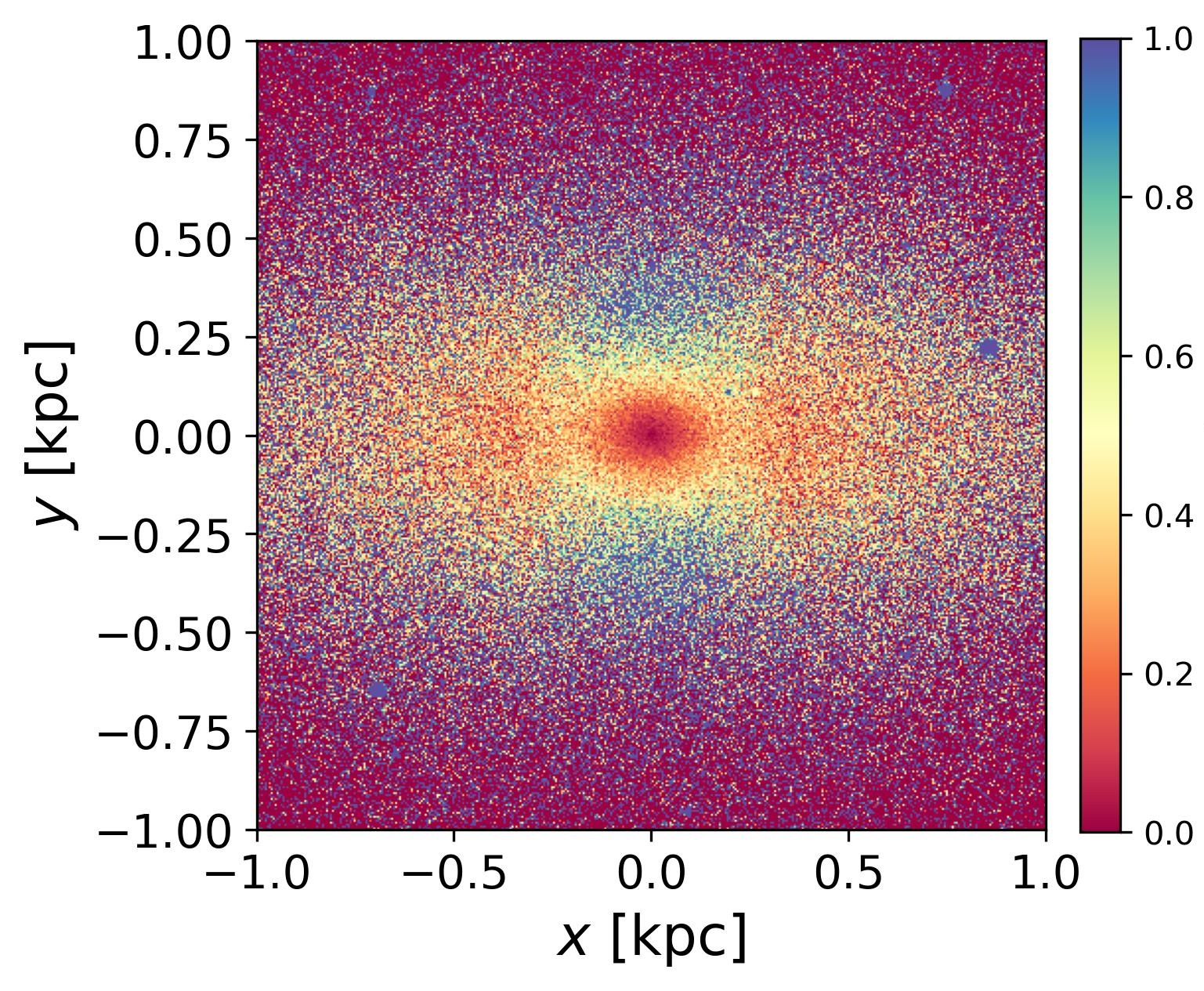} \\[14pt]

    \rotatebox{90}{\quad \quad \quad 4.0 Gyr} &
    \includegraphics[width=0.22\textwidth]{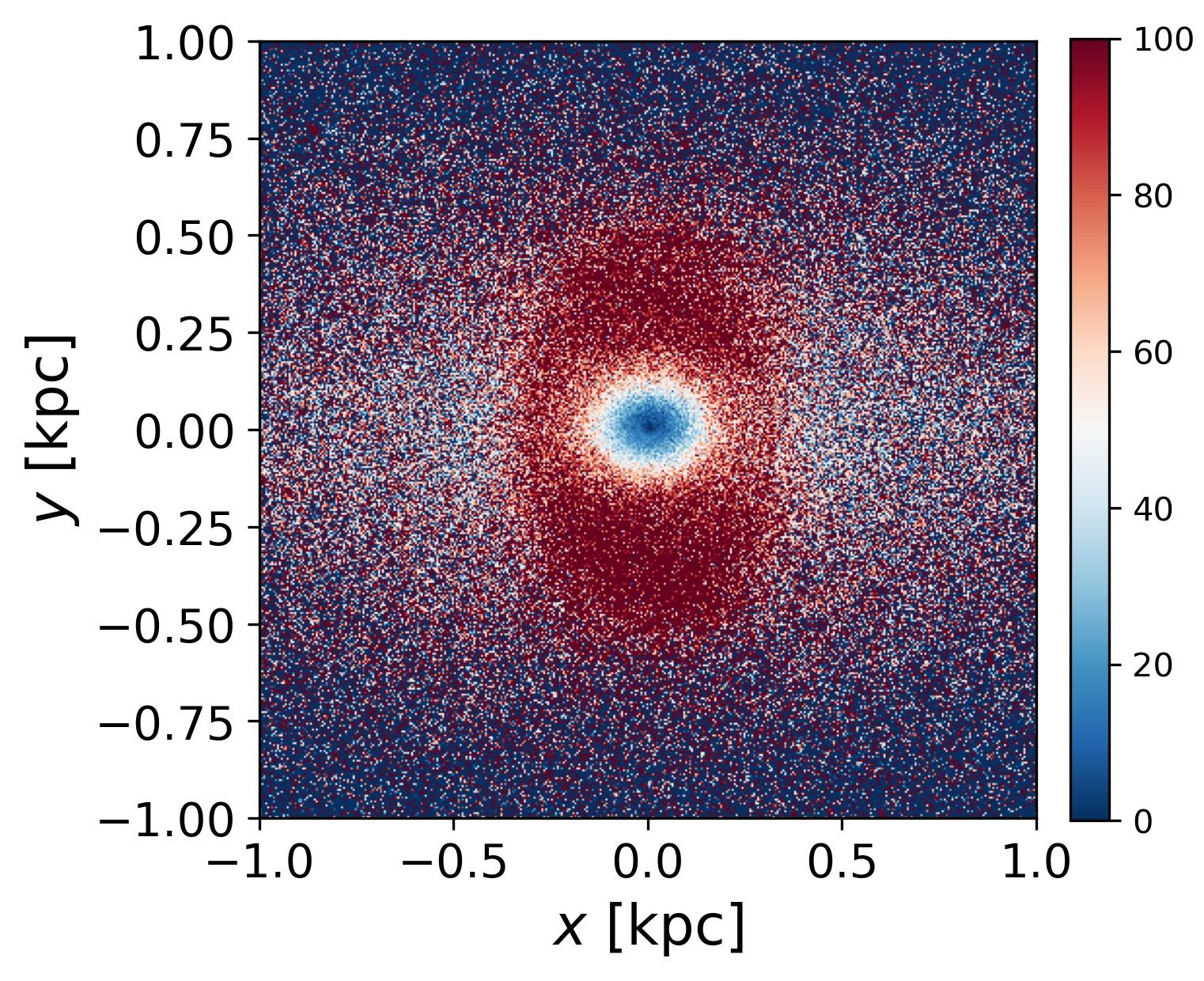} &
    \includegraphics[width=0.22\textwidth]{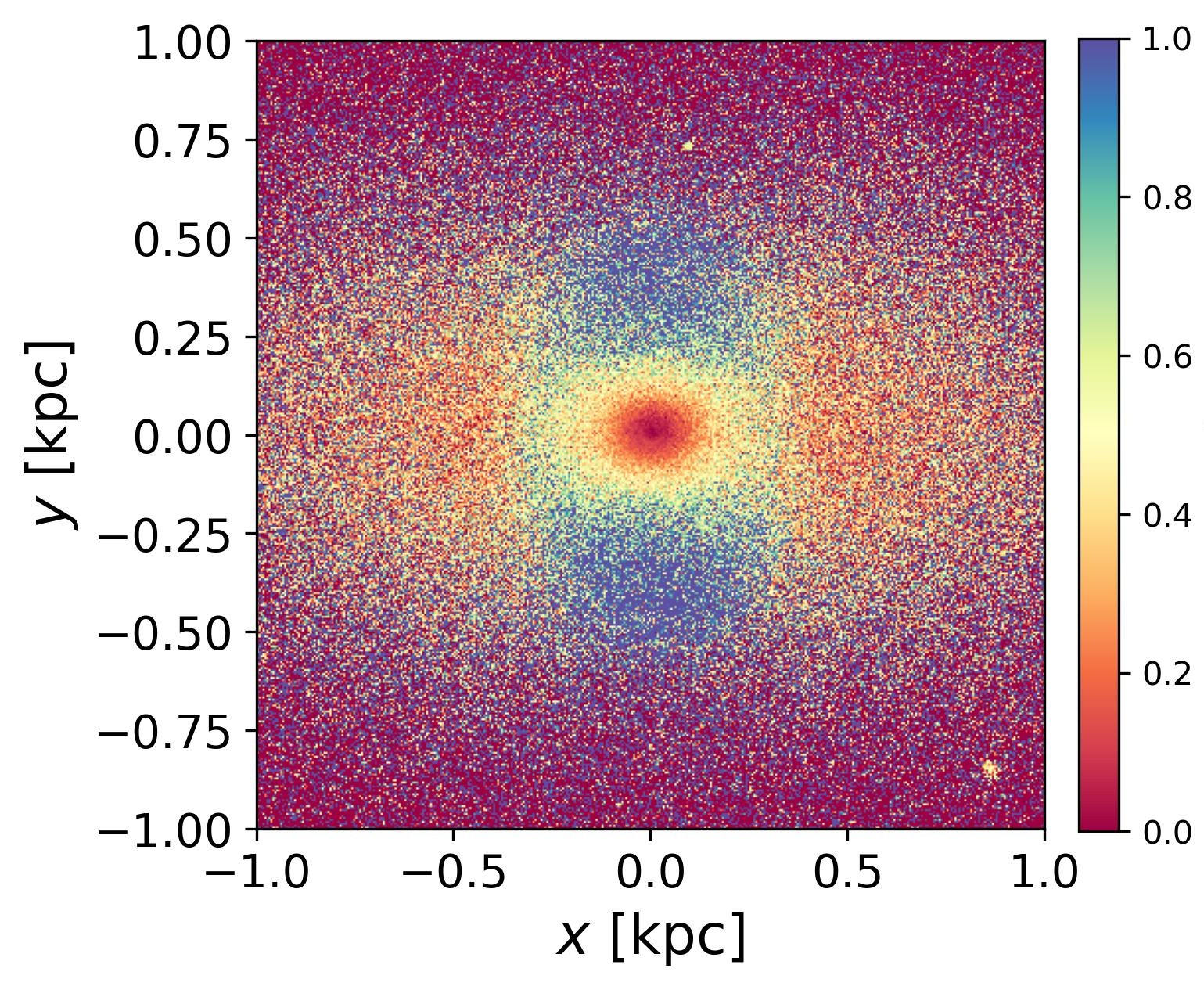} \\
  
    \end{tabular}
    \caption{Face-on projections of $v_{\phi}$ and $v_{\phi}/\sigma$ for all new stars without an age cut. The color bars are fixed from 0 to 100 km s$^{-1}$ for $v_{\phi}$ and from 0 to 1.0 for $v_{\phi}/\sigma$.}
    \label{fig:nsd_nocut}
\end{figure}

\begin{figure*}[t!]
    \centering
    \setlength{\tabcolsep}{0pt}
    \begin{tabular}{rcccc}

    & \textbf{SFT [Gyr]}  & \textbf{[Fe/H]} & \textbf{$v_{\phi}$ [km s$^{-1}$]} & \textbf{$v_{\phi}/\sigma$} \\[1pt]
   
    \rotatebox{90}{\quad \quad \quad 1.5 Gyr} &
    \includegraphics[width=0.22\textwidth]{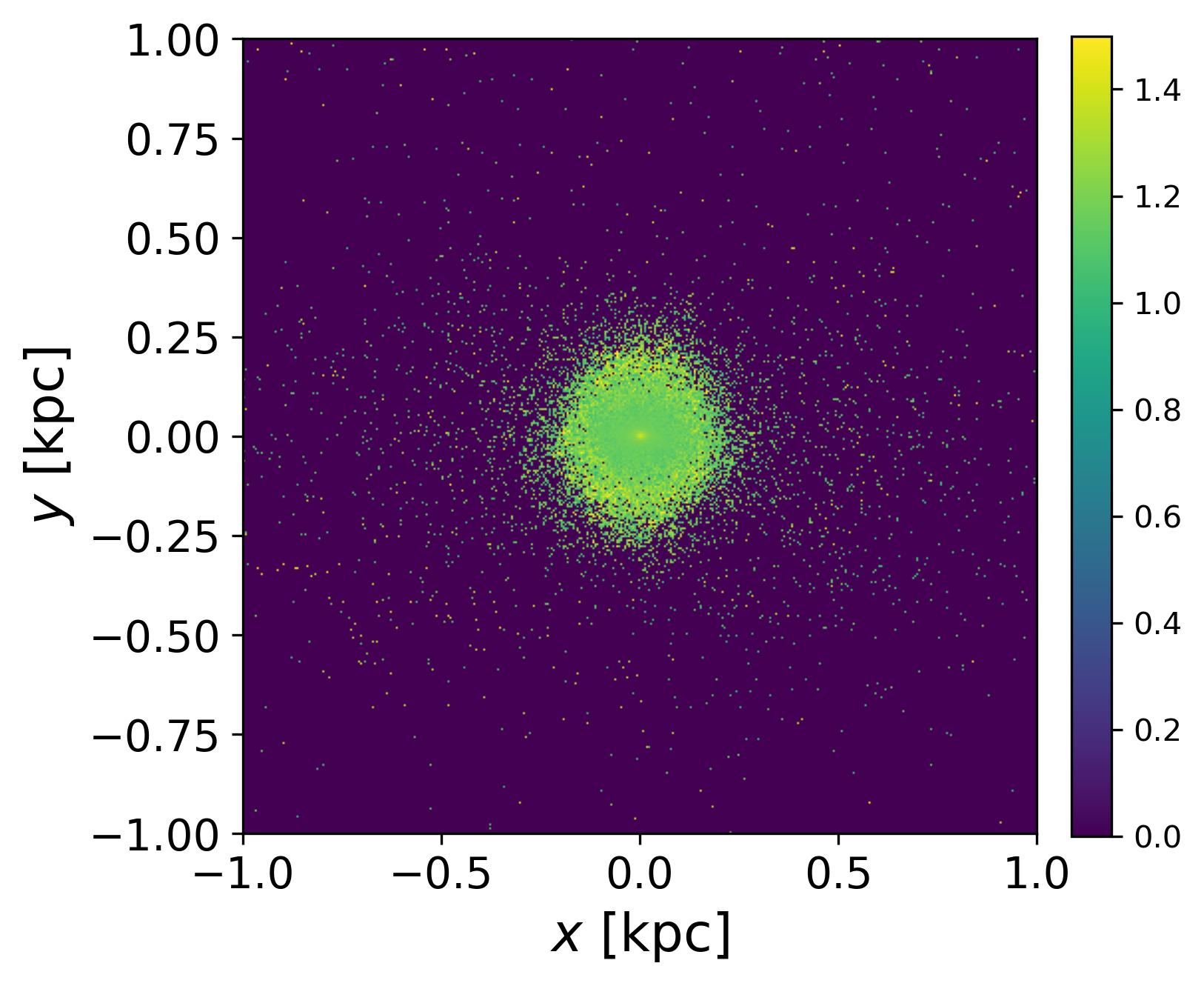} &
    \includegraphics[width=0.22\textwidth]{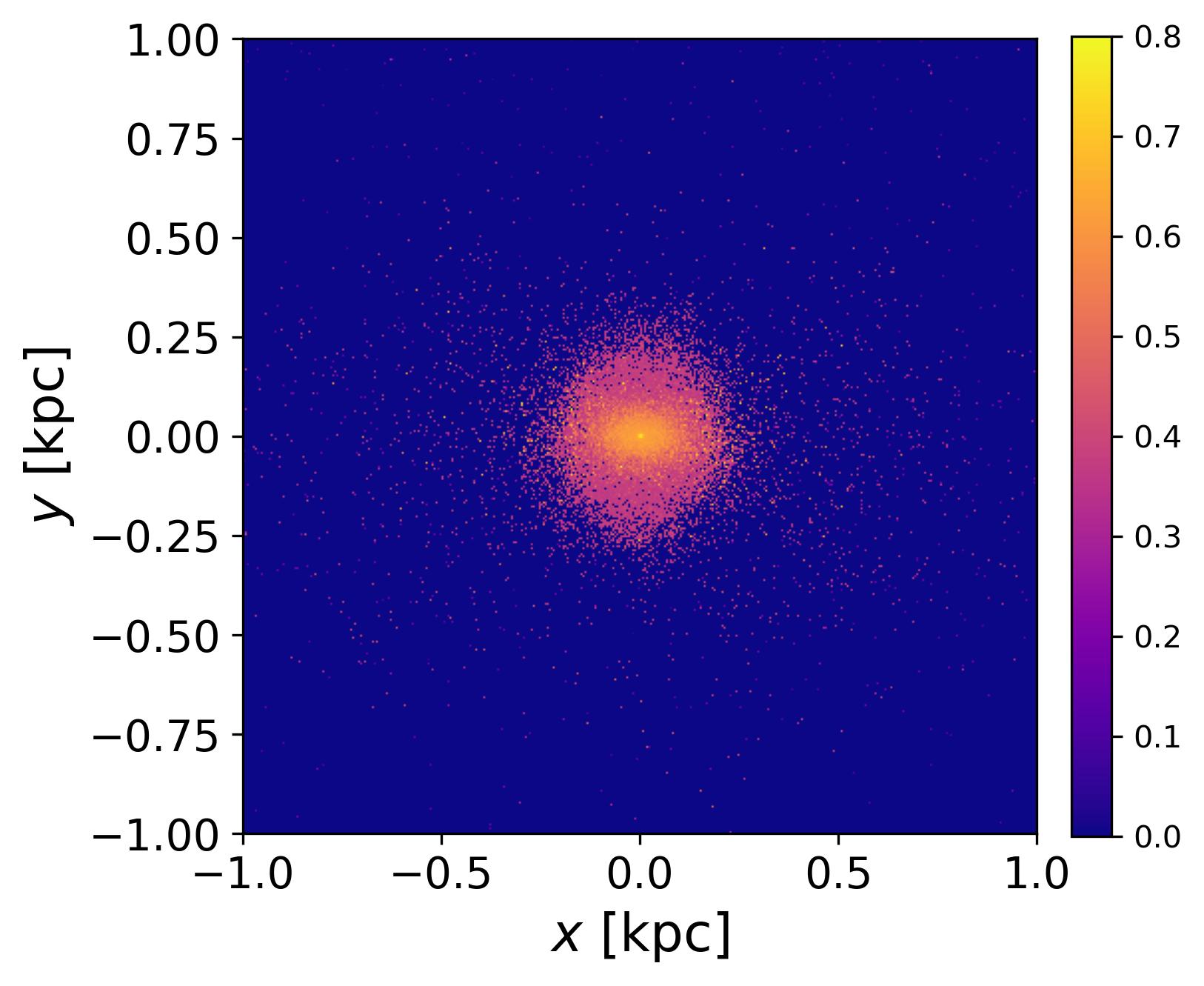} &
    \includegraphics[width=0.22\textwidth]{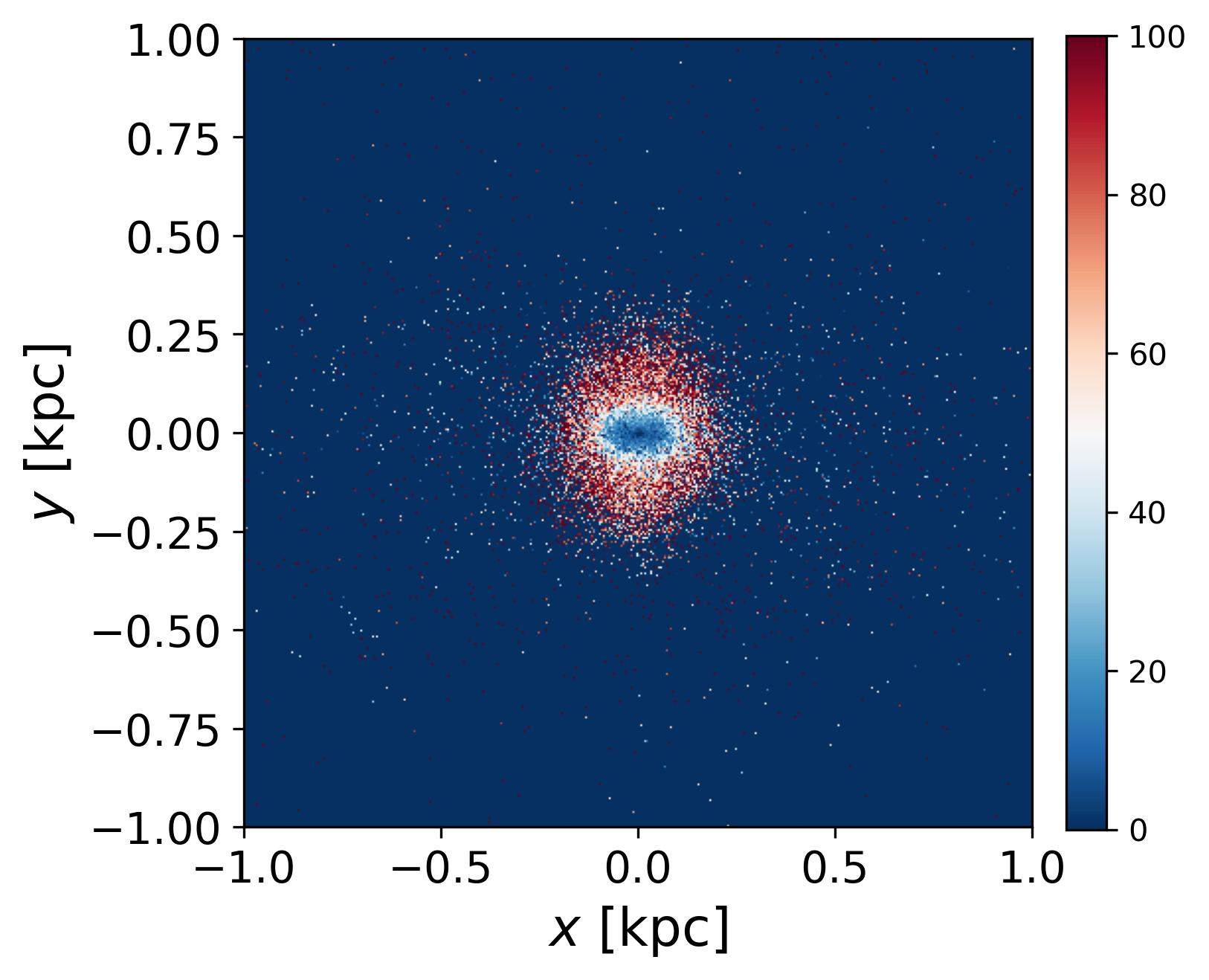} &
    \includegraphics[width=0.22\textwidth]{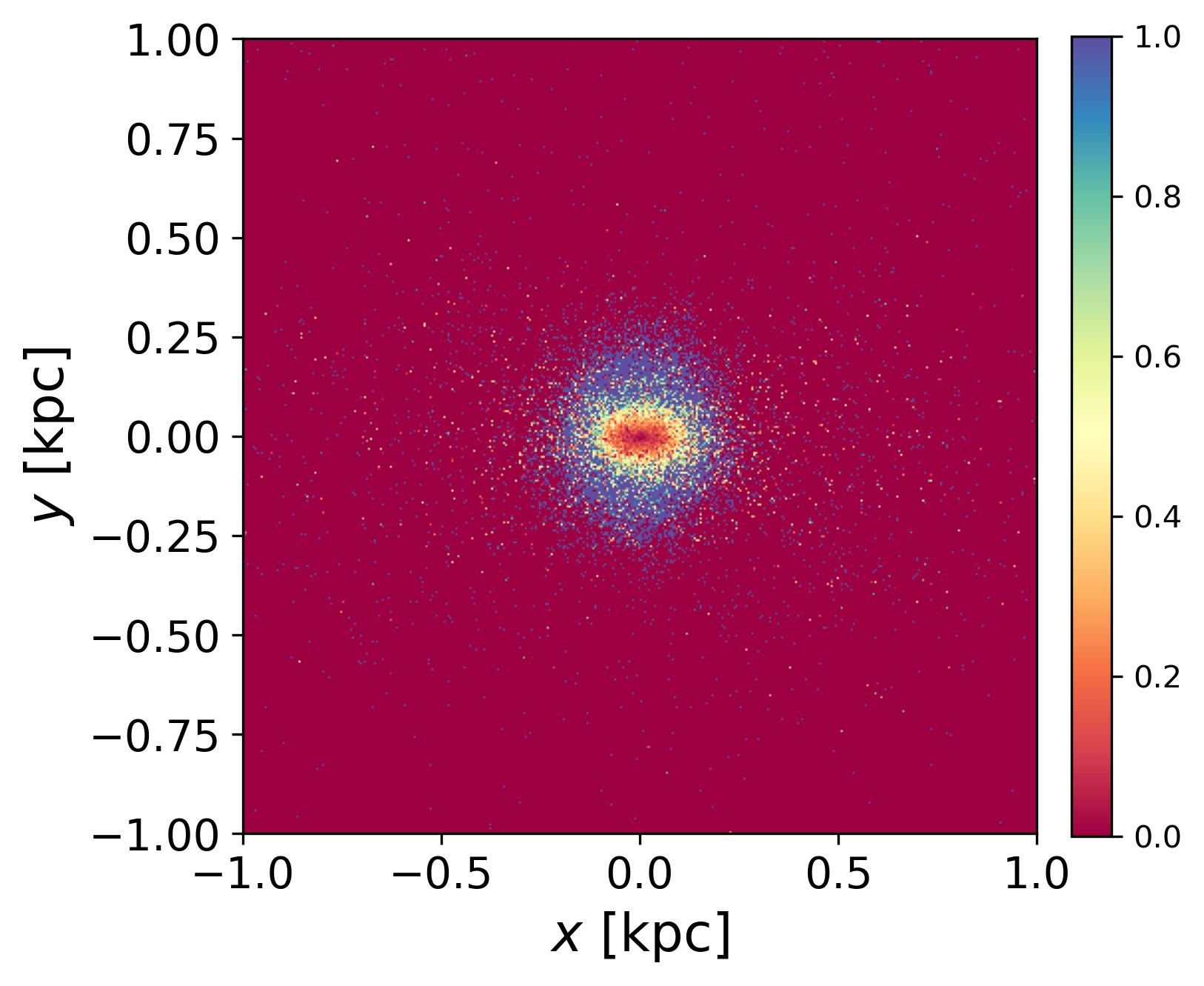} \\[1pt]
   
    \rotatebox{90}{\quad \quad \quad 2.6 Gyr} &
    \includegraphics[width=0.22\textwidth]{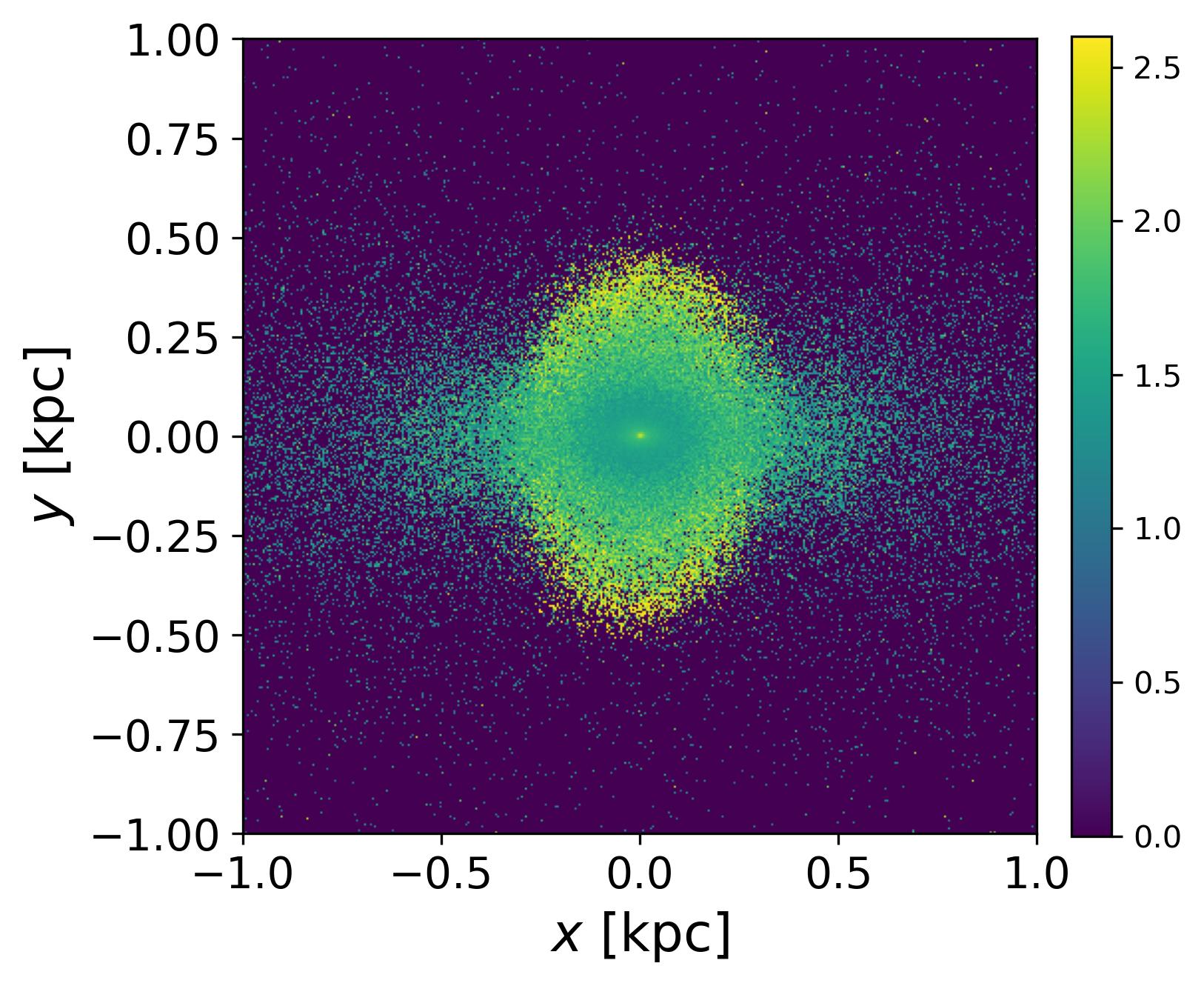} &
    \includegraphics[width=0.22\textwidth]{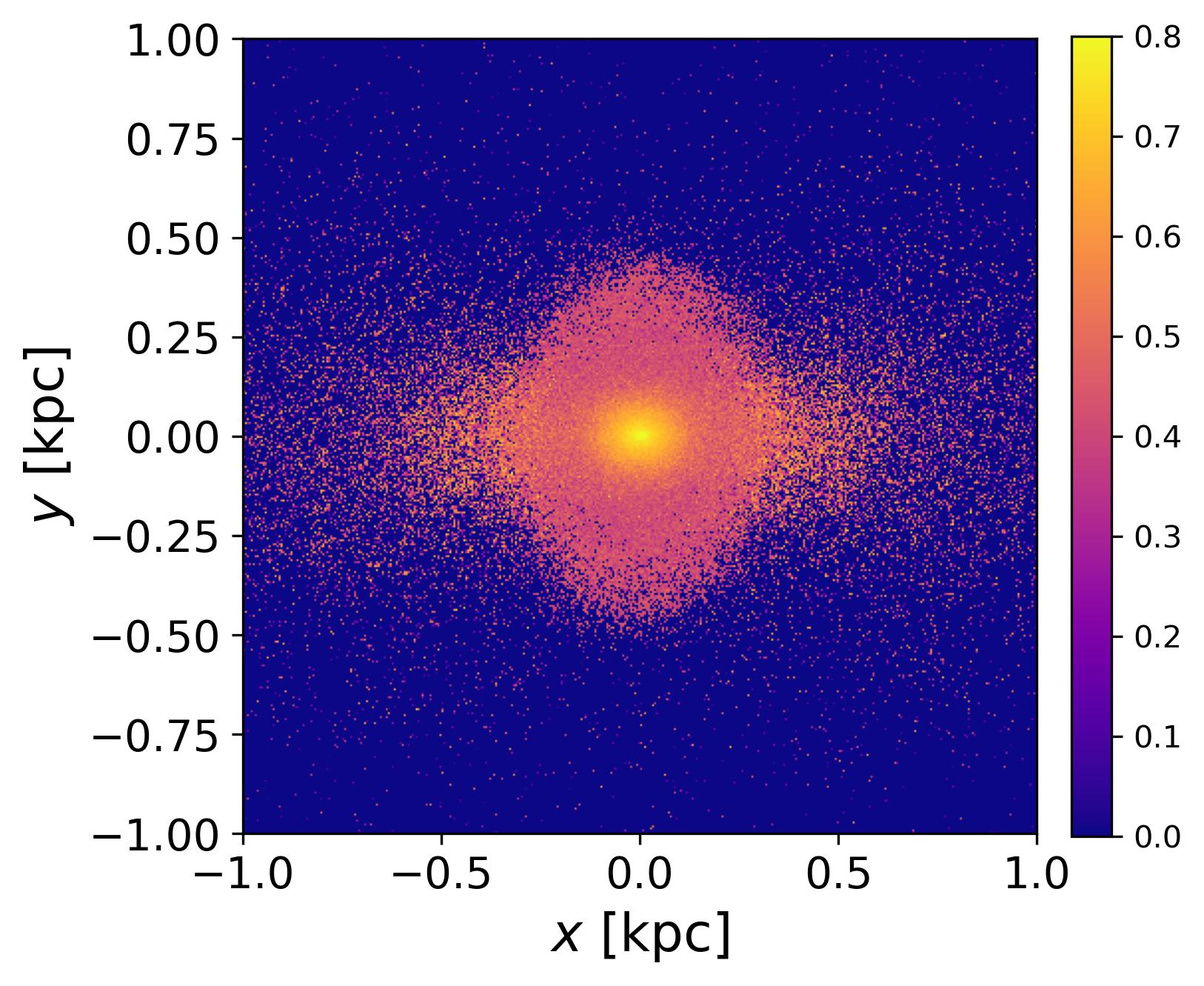} &
    \includegraphics[width=0.22\textwidth]{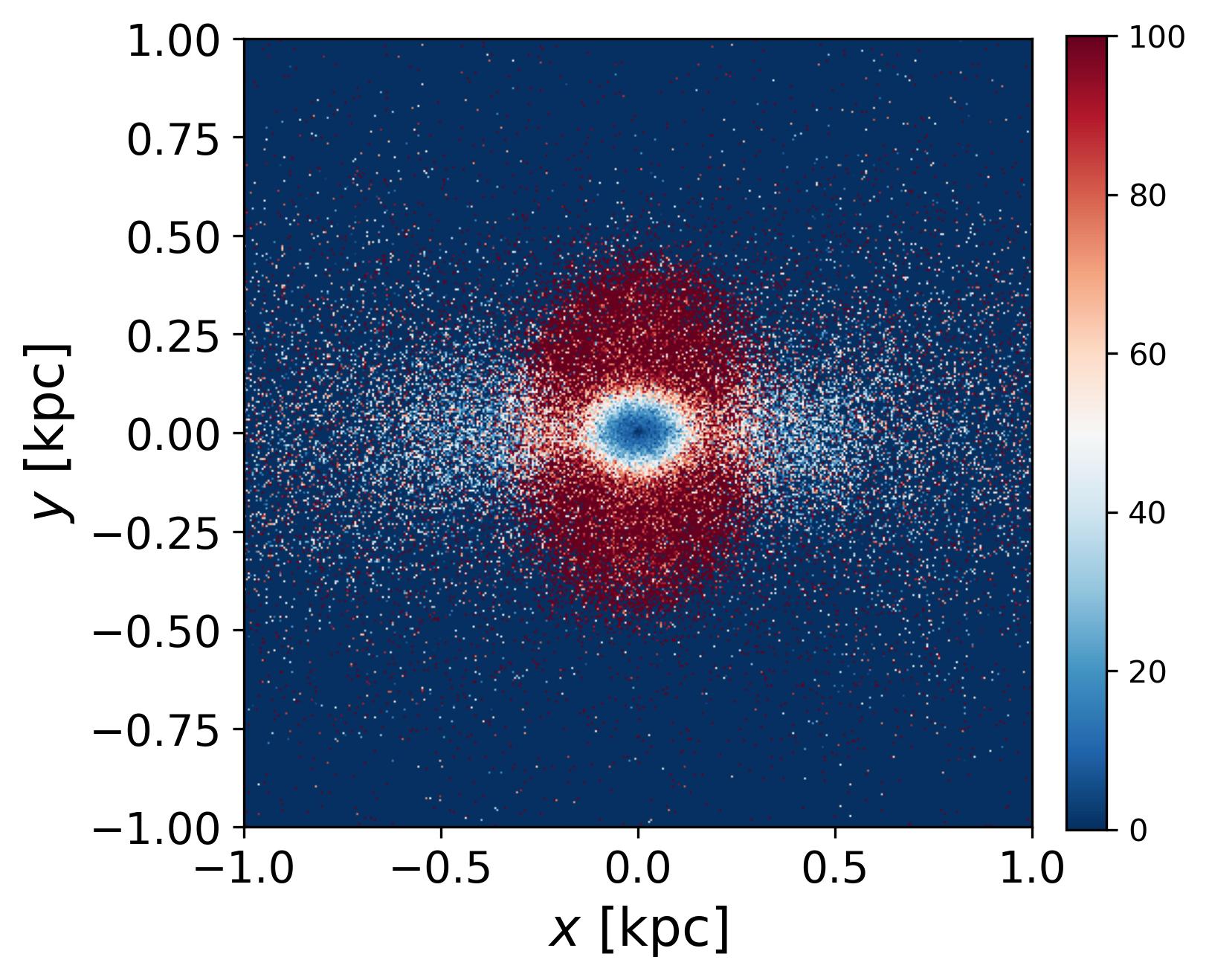} &
    \includegraphics[width=0.22\textwidth]{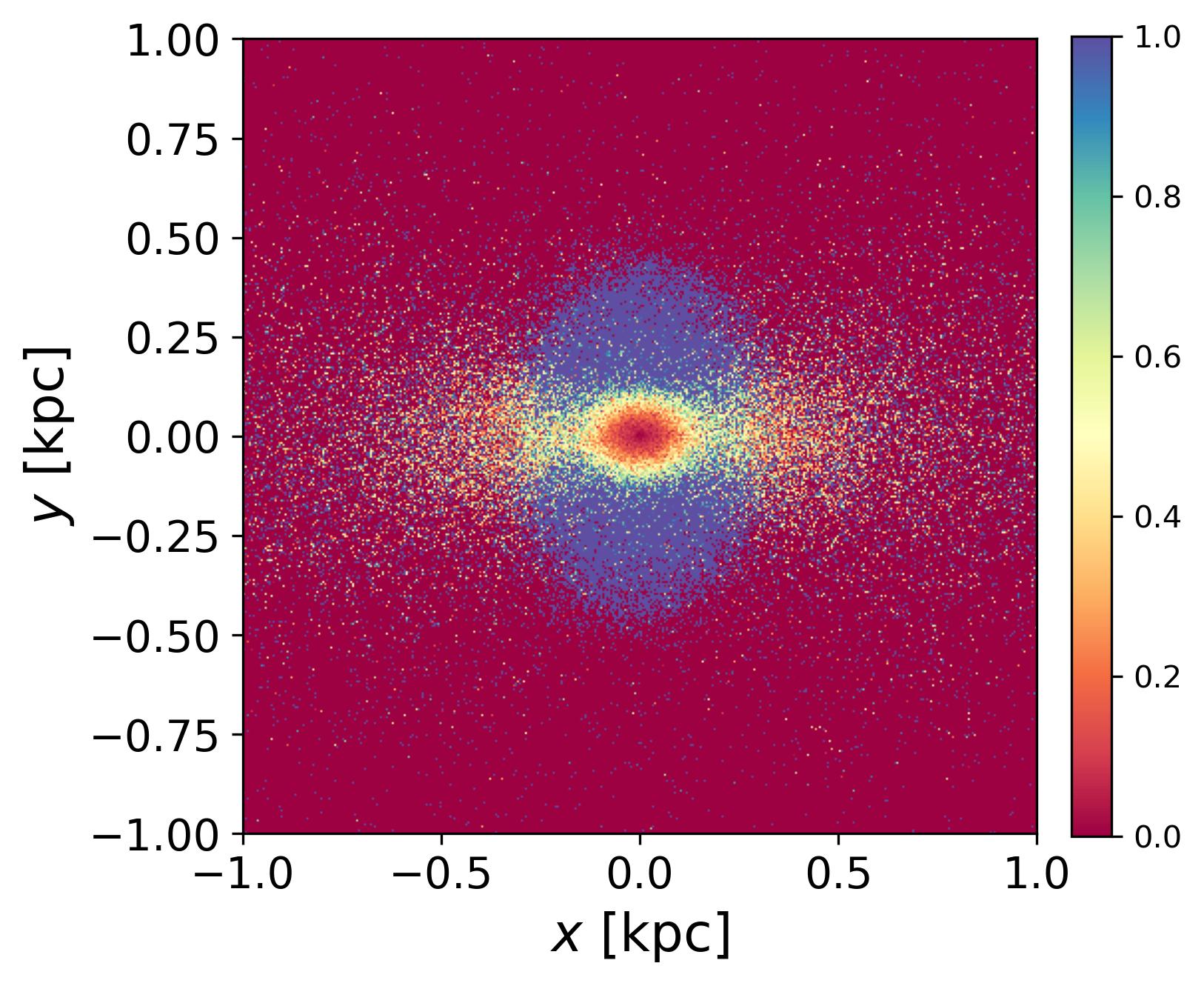} \\[1pt]
   
    \rotatebox{90}{\quad \quad \quad 4.0 Gyr} &
    \includegraphics[width=0.22\textwidth]{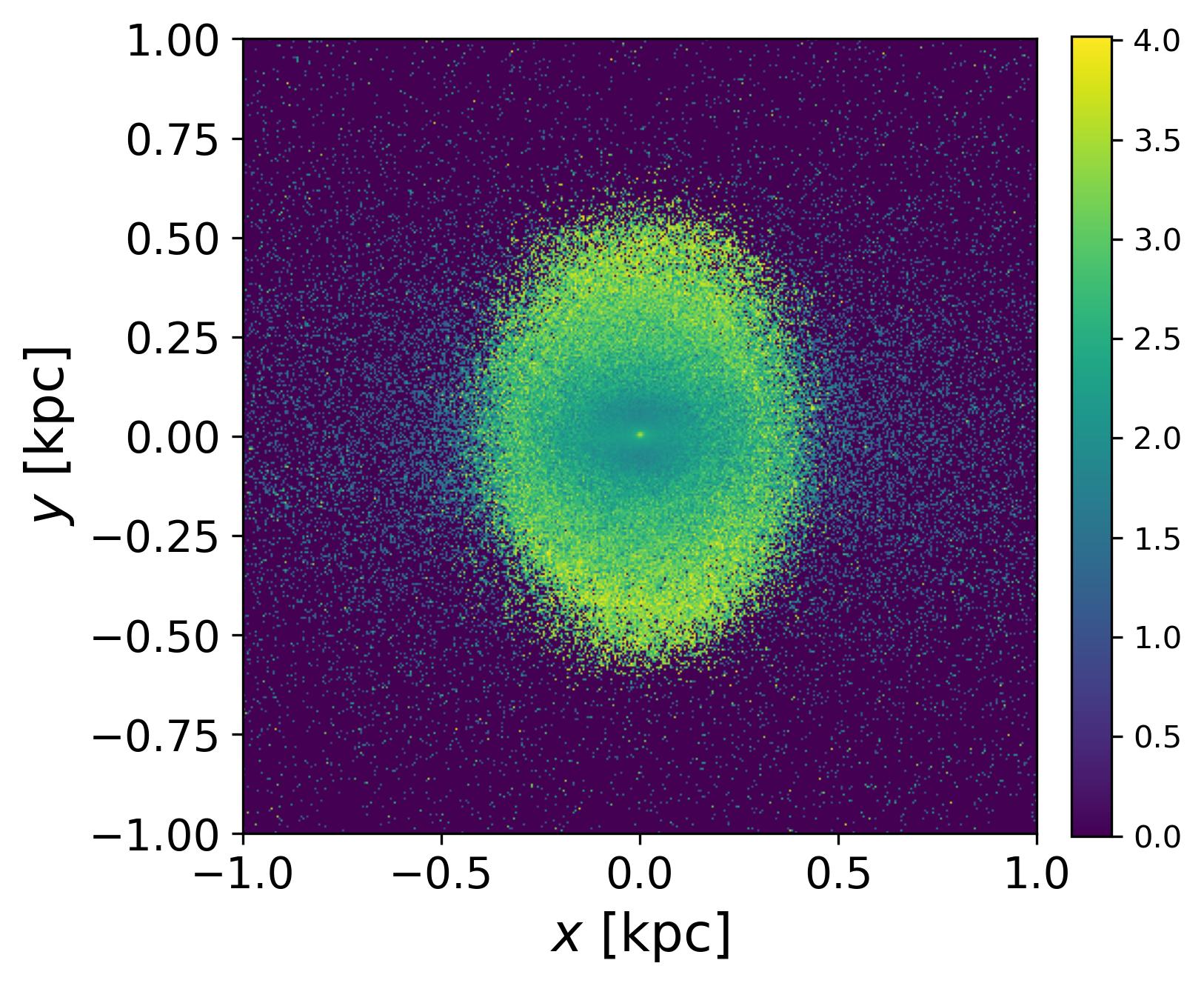} &
    \includegraphics[width=0.22\textwidth]{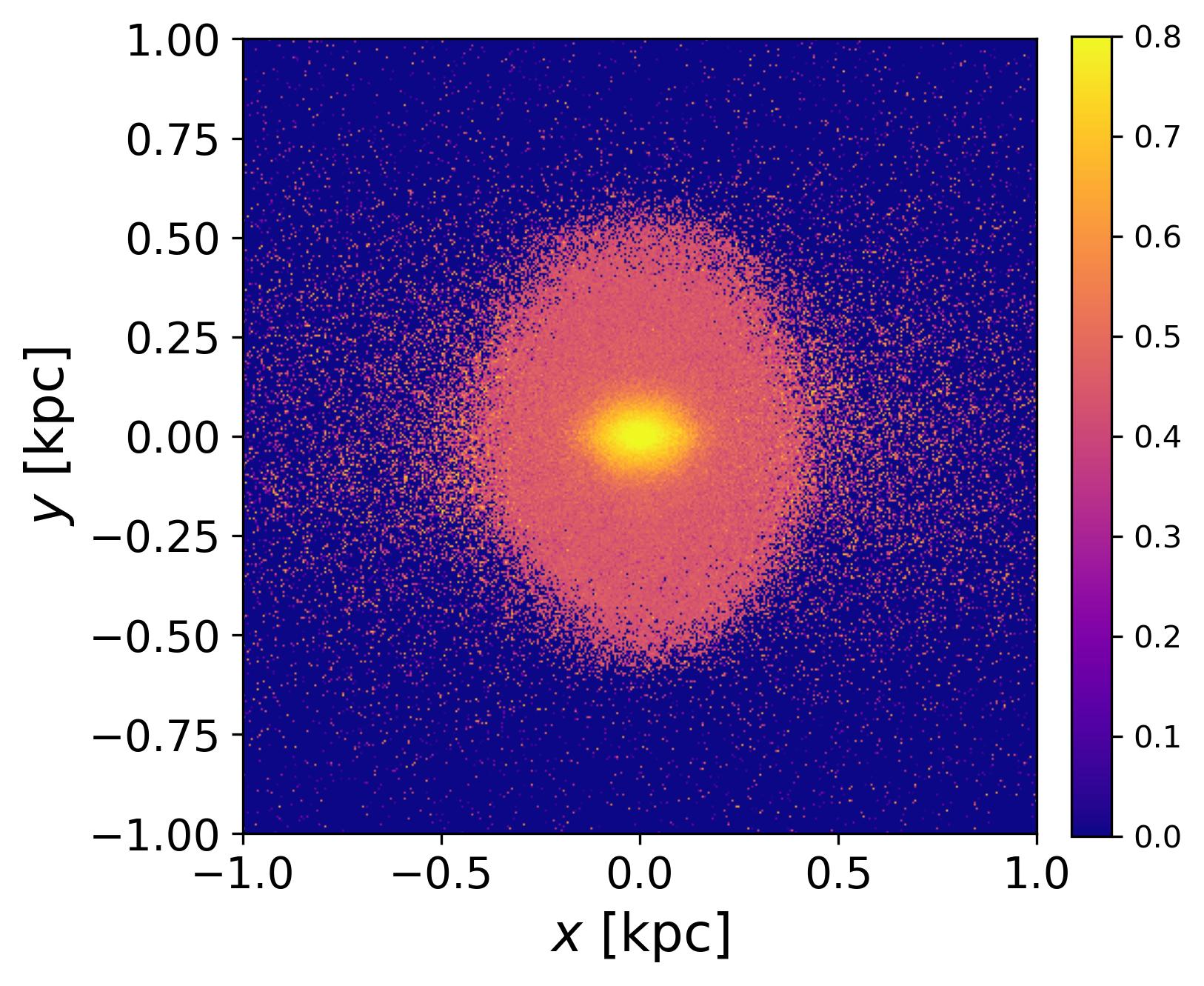} &
    \includegraphics[width=0.22\textwidth]{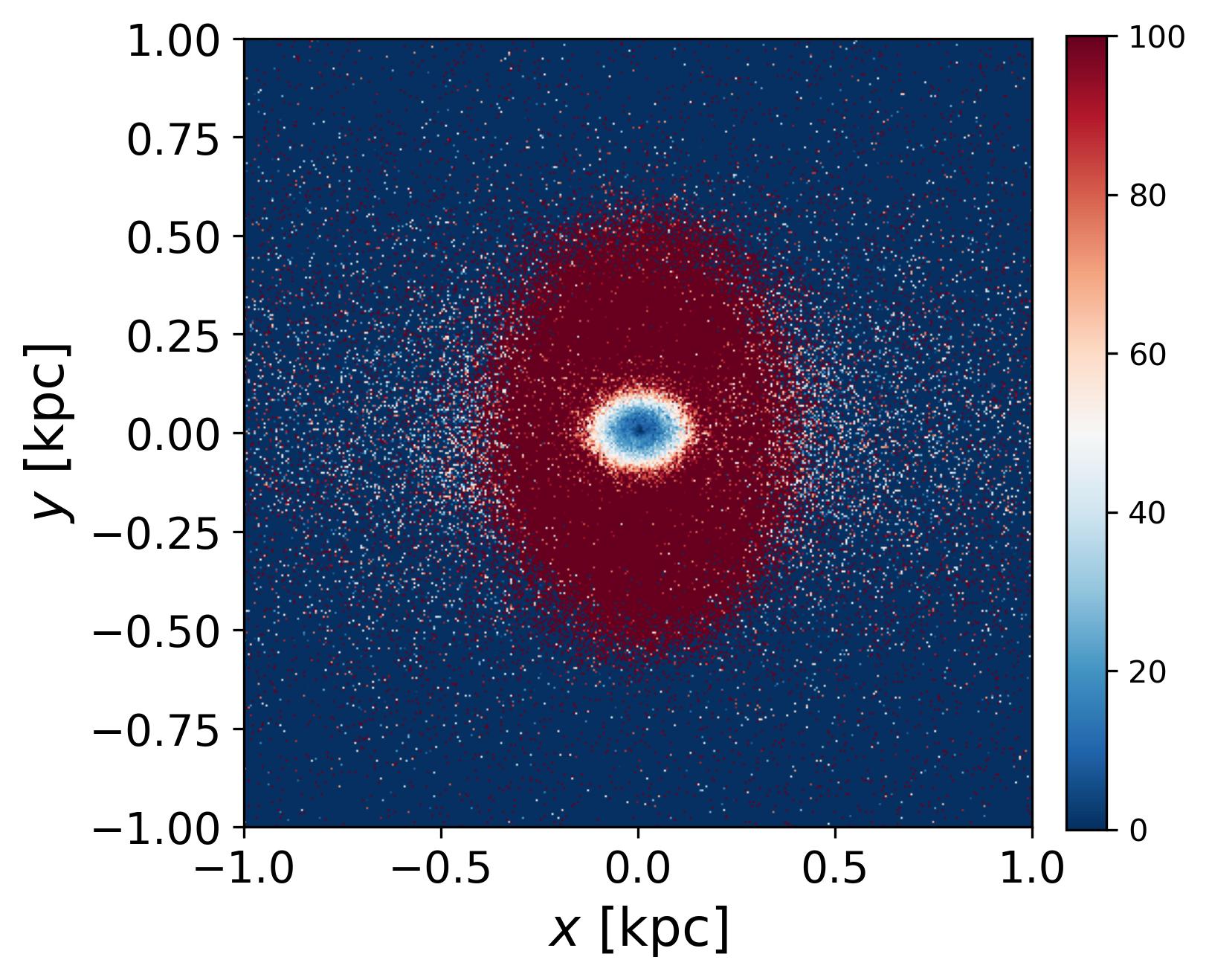} &
    \includegraphics[width=0.22\textwidth]{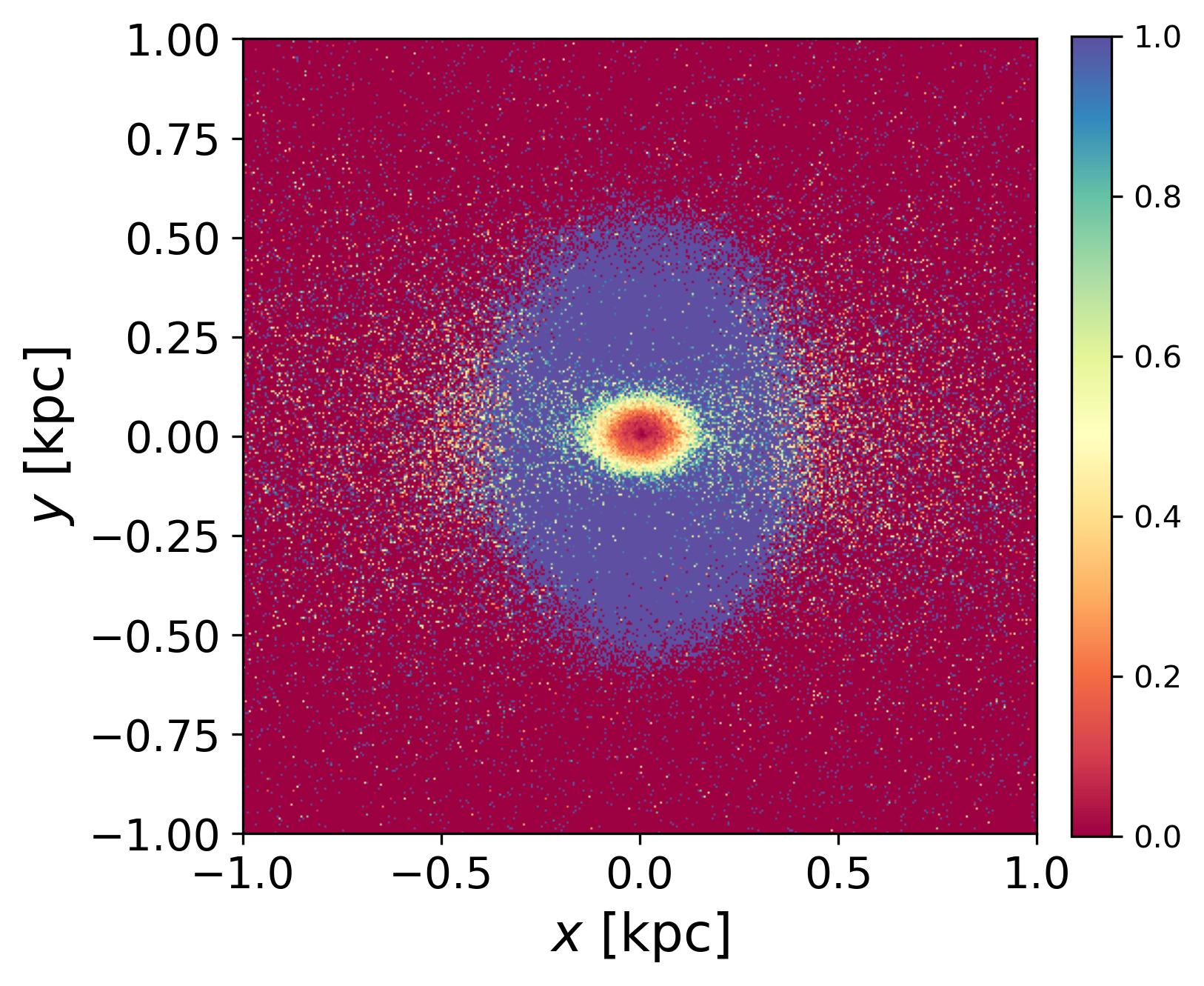} \\
   
    \end{tabular}
    \caption{Face-on projections of SFT, [Fe/H], $v_{\phi}$, and $v_{\phi}/\sigma$ for the new stars born after the bar formation, $t_{\rm bar}=1 \ \Gyr$. The color bar for SFT is fixed from 0 to the selected time. The color bars are fixed from 0 to 0.8 for [Fe/H], from 0 to 100 km s$^{-1}$ for $v_{\phi}$, and from 0 to 1.0 for $v_{\phi}/\sigma$.}
    \label{fig:nsd_agecut}
\end{figure*}

\begin{figure*}[t!]
    \centering
    \includegraphics[width=0.32\textwidth]{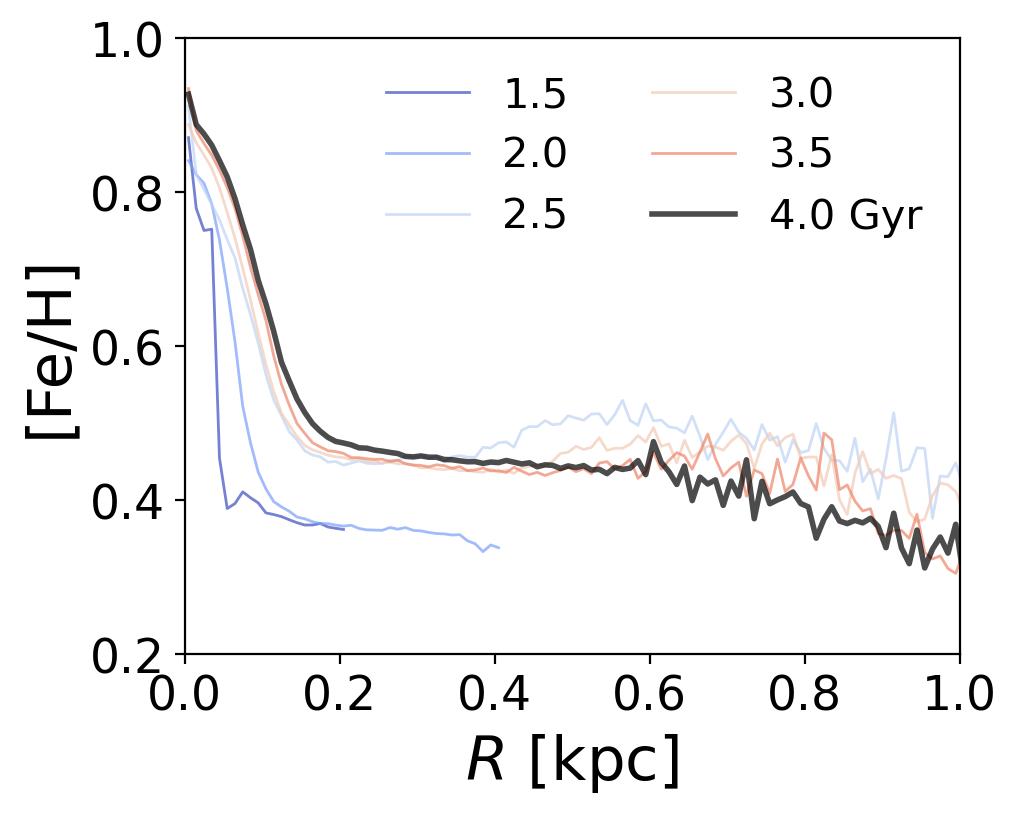}
    \includegraphics[width=0.32\textwidth]{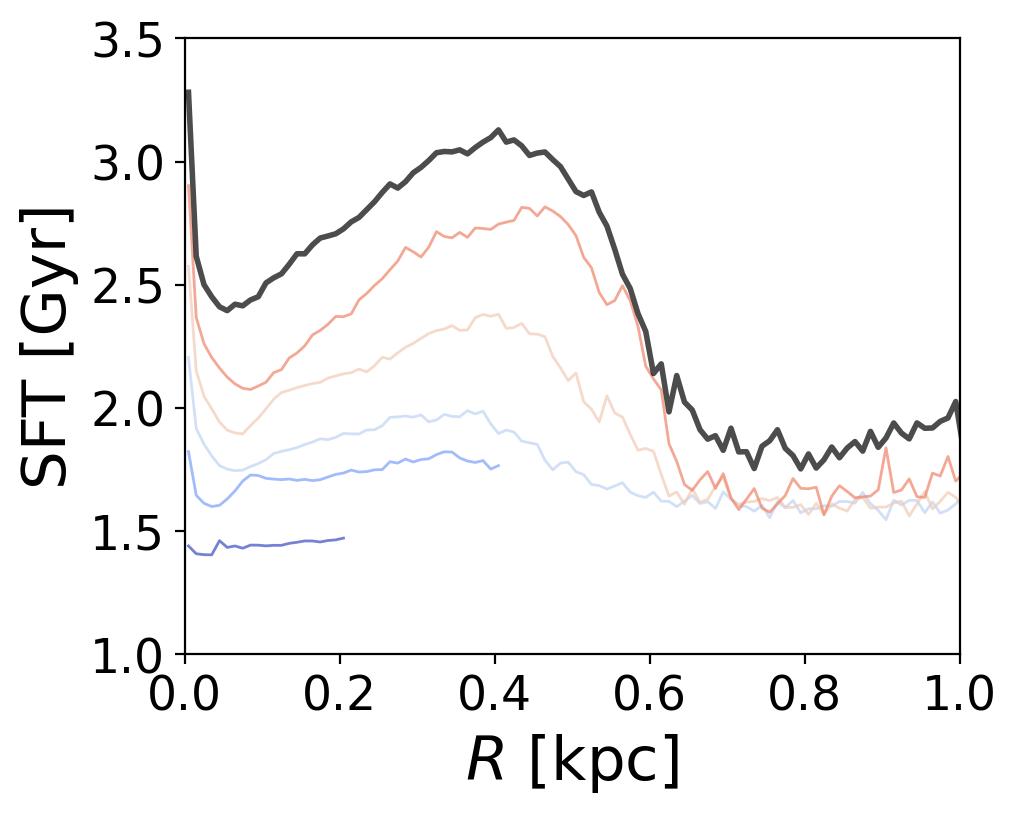}

    \includegraphics[width=0.32\textwidth]{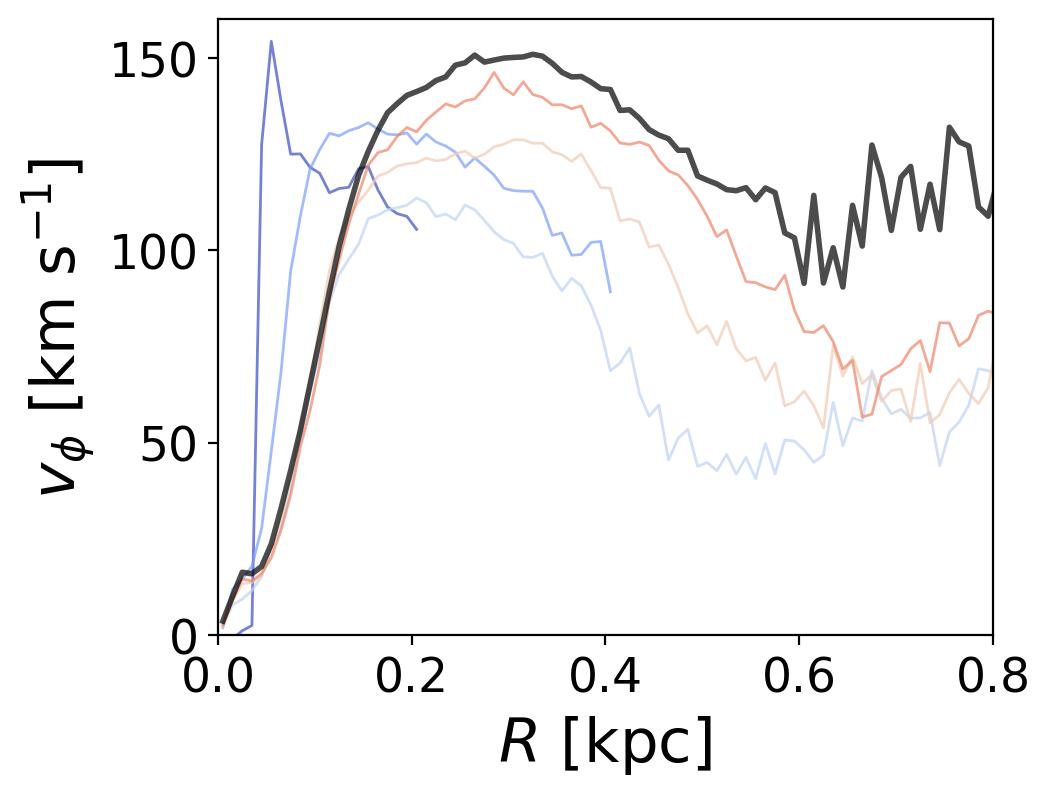}
    \includegraphics[width=0.32\textwidth]{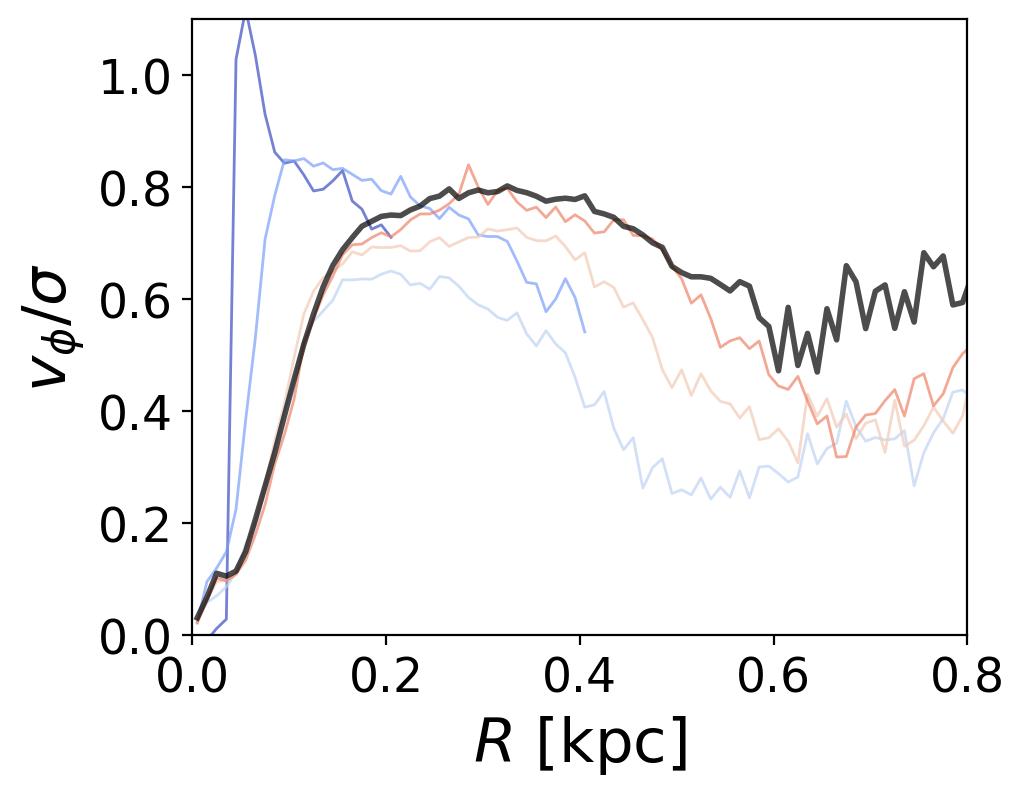}
    
    \caption{Radial profiles of properties within 0.8 kpc at six epochs from 1.5 to 4.0 Gyr, spaced at 0.5 Gyr intervals. Colors change from blue to red with increasing time, and the black line shows the profile at 4.0 Gyr. Metallicity, star formation time (SFT), $v_{\phi}$, and $v_{\phi}/\sigma$ are calculated for the new stars formed after the bar formation, $t_{\rm bar}=1\,\mathrm{Gyr}$.}
    \label{fig:profiles}
\end{figure*}

\subsection{Fourier analysis and bar strength}\label{appendix:fourier}

To quantify the bar strength, we perform a Fourier analysis of the mass distribution:
\begin{equation}\label{eq:f2}
F_{m}(R) = \frac{\sum_{j} \mu_{j} \, e^{i m \phi_{j}}}{\sum_{j} \mu_{j}},
\end{equation}
where \(\mu_{j}\) and \(\phi_{j}\) are the mass and azimuthal angle, respectively, of the \(j\)th particle in a radial annulus of width \(\Delta R = 0.01\,\mathrm{kpc}\), and \(m\) is the azimuthal wavenumber (multipole order). We focus on the \(m=2\) mode, which traces the bar strength within 8 kpc.

Figure~\ref{fig:map8kpc} shows the temporal evolution of the radial distribution of the $m=2$ Fourier mode (Eq.~\ref{eq:f2}) and the gas mass within 8 kpc, using a bin size of 0.01 kpc. We fix the color bar from 0 to 0.2, with red indicating the bar region ($F_2>0.2$). Owing to the enhanced mass and force resolution relative to model r1c14b05 in Paper~I (in which the bar region reaches only 3.7 kpc with $F_2>0.15$), the bar in model r1c14b05 grows stronger and longer, reaching around 5 kpc ($F_2>0.2$) due to the reduced numerical noise \citep{kwak26a}. The elongation of the bar is clearly reflected in the gas distribution, as the bar exerts perturbation that drives gas inflows toward the center (\citealt{combes85, combes93, athanassoula92, seo19}; Paper I).

\subsection{Age selection and radial profiles}\label{appendix:agecut_profile}
To examine the properties of the nuclear structures, we apply an additional age criterion and select only new stars born after bar formation at $  t_{\rm bar}=1  $, following the age-dating method of \citep{baba20,sanders24,desafreitas23a}.
For comparison, Fig.~\ref{fig:nsd_nocut} shows the face-on projections of $  v_{\phi}  $ and $  v_{\phi}/\sigma  $ at 1.5, 2.6, and 4.0 Gyr without any age selection on the new stars.
Figure~\ref{fig:nsd_agecut} presents metallicity [Fe/H], SFT, $ v_{\phi} $, and $ v_{\phi}/\sigma $ after the age selection. Figure~\ref{fig:profiles} shows the radial profiles of those values from 1.5 to 4.0 Gyr at 0.5 Gyr intervals. This age-selection clearly distinguishes the nuclear structures and reveals the inside-out trend from early times to the end of the evolution. With our definition of the NSC as $  v_{\phi}/\sigma < 0.3  $, the red region corresponds to the NSC while the bluish region corresponds to the NSD (see $v_{\phi}/\sigma$ in Fig.~\ref{fig:nsd_agecut}). As shown in the radial profiles of $  v_{\phi}/\sigma  $ (Fig.~\ref{fig:profiles}), changing the NSC definition to $  v_{\phi}/\sigma < 0.5  $ does not significantly change the NSC size. In the $  v_{\phi}  $ map (Fig.~\ref{fig:nsd_agecut}), the NSD and NSC exhibit distinct kinematics, with the NSC being concentrated at the center, consistent with \cite{neumayer20}. At 1.5 Gyr, the small NSD and NSC form and are cleanly isolated from unrelated stars. During this inside-out formation, the NSC remains compact while the NSD expands over time. The same patterns appear in the face-on projections of SFT and metalicity [Fe/H] in Fig.~\ref{fig:nsd_agecut}. The color bar for the SFT map is fixed from 0 to the time of each panel, with yellower colors indicating younger stellar populations. As the bar funnels gas from the outer disk to the NSD edge, the NSD grows in size and expands over time, so the outer edge of the NSD shows the youngest stellar populations. The NSC also exhibits signs of star formation, though this is very confined to the center. Because the NSC remains very compact, its stellar populations appear mixed unlike yellow region seen at the NSD edge. 

\subsection{Accretion of the star cluster into the NSC}\label{appendix:starcluster}
\begin{figure}[t!]
    \centering
    \includegraphics[width=0.5\textwidth]{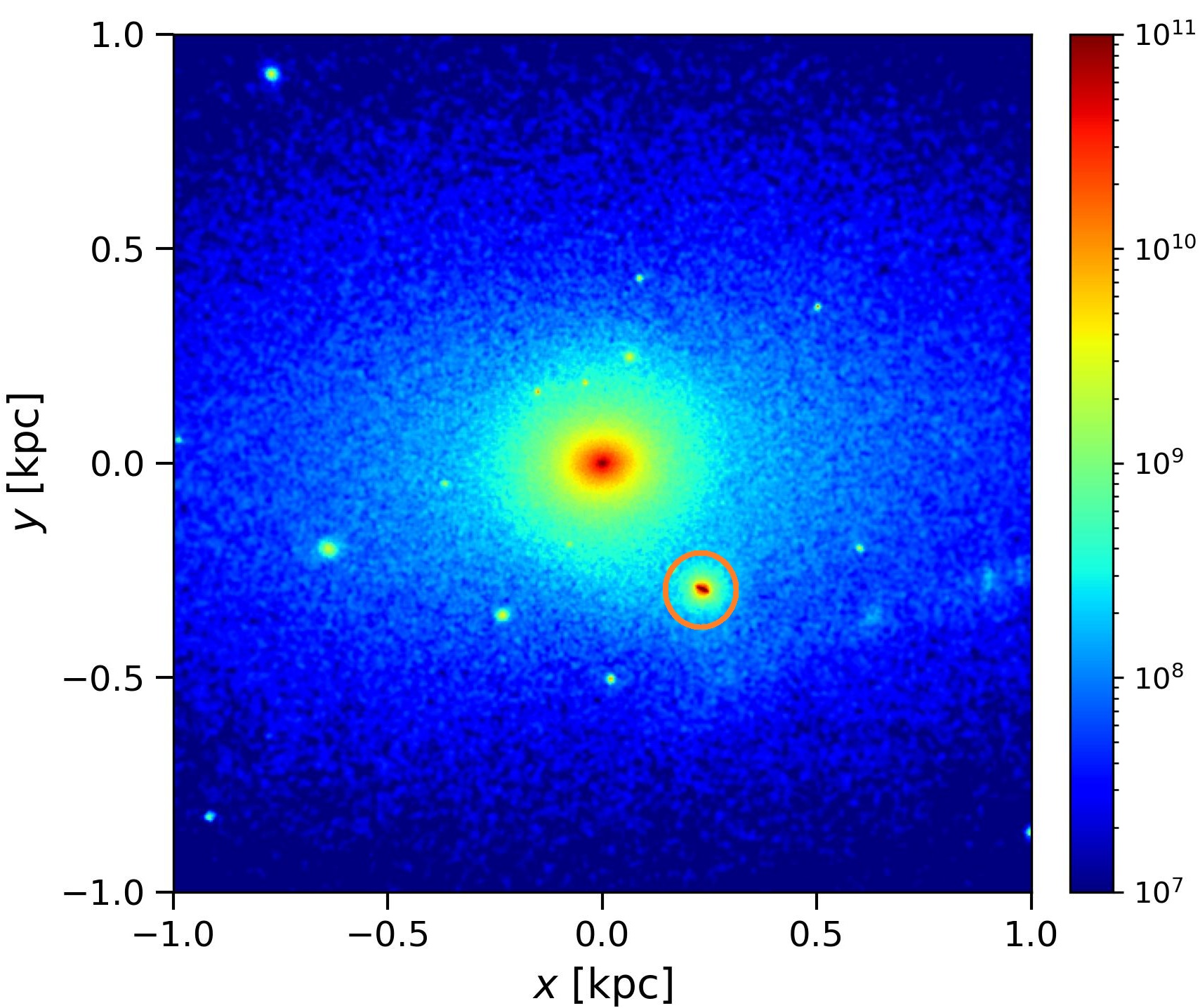}    
    \includegraphics[width=0.5\textwidth]{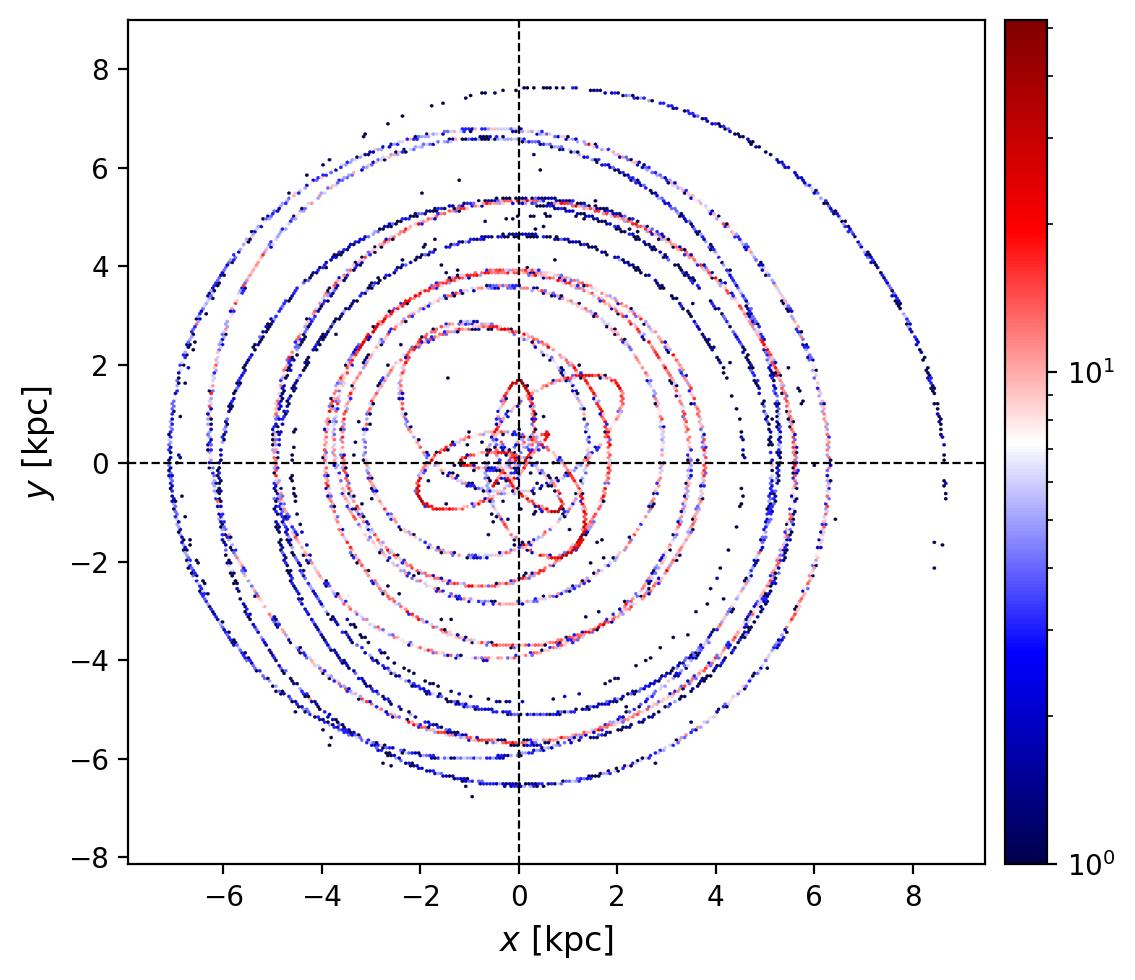}    
    \caption{\textit{Top}: Face-on projection of the surface density distribution of the new stars within 1 kpc at 2.0 Gyr before the merger with the massive star cluster (orange circle). The color bar is in units of $\Sigma_{\star} \ \rm kpc^{-2}$. \textit{Bottom}: Density distribution of the birth positions of the bound stars in the accreted massive star cluster. The color bar indicates the number of stars born in each bin.}  
    \label{fig:accretion}
\end{figure}

To understand the star cluster population and their infalling history, we applied the FOF algorithm (e.g., \citealt{davis85}) to identify bound star clusters in our galaxy model at 2 Gyr, which is before the merger event takes place at 2.1 Gyr (see Fig.~\ref{fig:nscnsd}). The moment prior to the merger event at 2 Gyr is visualized in the top panel of Fig.~\ref{fig:accretion}, where the central object is the NSC. We find three clusters with masses of approximately $10^{7}\Msun$, 13 clusters between $10^{6}$ and $10^{7}\Msun$, and about 200 clusters between $10^{5}$ and $10^{6}\Msun$. The effects of minor mergers do not appear to be significant in the mass growth (top-right panel of Fig.~\ref{fig:nscnsd}), which is consistent with the observational results that merger events play a more dominant role in the growth of NSCs in dwarf galaxies \citep{fahrion22, fahrion22b, fahrion24}. The accreted massive star cluster at 2.1 Gyr with a mass of $3\times10^{7} \Msun$ is the most massive one among the cluster population (orange circle in the top panel of Fig~\ref{fig:accretion}). The density distribution of the birth positions of the newborn stars that are bound to the accreted star cluster is illustrated in the bottom panel of Fig.~\ref{fig:accretion}, with the red color in the color bar indicating higher star formation and mass growth. Apparently, the star cluster first forms in the outer region of the disk. It orbits around the disk, and its orbit slowly decays. As it reaches the bar region (4 to 5 kpc), the density of newborn stars increases, and this agrees with the fact that the bar promotes the star cluster to become more massive \citep{ali23}. After the accreted star cluster's orbit decays further inward, its orbit is perturbed by the bar and then falls deeper into the nuclear region on elongated orbits, which is also evident in the oscillations of the mass of the NSD in the top-right panel of Fig.~\ref{fig:nscnsd}.

\end{appendix}

\end{document}